\documentclass[aps,prb,twocolumn,groupaddress]{revtex4-1}

\usepackage{amssymb,amsmath}
\usepackage[varg]{txfonts}
\usepackage[usenames,dvipsnames]{color}
\usepackage[colorlinks,linkcolor=blue,citecolor=blue,urlcolor=blue]{hyperref}
\usepackage{graphicx}
\usepackage[utf8]{inputenc}
\usepackage[T1]{fontenc}

\newcommand{\dd}{\mathrm{d}}

\newcommand{\iu}{\mathrm{i}}

\mathchardef\mhyphen="2D

\begin{document}

\title{Time-dependent many-body treatment of electron-boson dynamics:
  application to plasmon-accompanied photoemission}

\author{M. Sch{\"u}ler}\email[]{michael.schueler@physik.uni-halle.de} 
\author{J. Berakdar}
\author{Y. Pavlyukh} 
\affiliation{Institut
  f\"{u}r Physik, Martin-Luther-Universit\"{a}t Halle-Wittenberg,
  06099 Halle, Germany}
\date{\today}
\begin{abstract}
  Recent experiments access the time-resolved photoelectron signal
  originating from plasmon satellites in correlated materials and 
  address their build-up and decay in real time. Motivated by
  these developments, we present the Kadanoff-Baym formalism for the
  nonequilibrium time evolution of interacting fermions and
  bosons. In contrast to the fermionic case the bosons are described by
   second-order differential equations.  Solution of the bosonic
  Kadanoff-Baym equations\,---\,which is the central ingredient of
  this work\,---\,requires substantial modification of the usual
  two-times electronic propagation scheme. The solution is quite
  general and can be applied to a number of problems, such as the
  interaction of electrons with quantized photons, phonons and other
  bosonic excitations. Here, the formalism is applied to the
  photoemission from a deep core hole accompanied by plasmon
  excitation. We compute the time-resolved photoelectron spectra and
  discuss the effects of \emph{intrinsic} and \emph{extrinsic}
  electron energy losses and their interference.
\end{abstract}

\maketitle

\section{Introduction}

The impressive advances in the field of attosecond
metrology~\cite{kling_attosecond_2008,krausz_attosecond_2009,
  nagele_time-resolved_2011,pazourek_time-resolved_2013} lead to new
insights into the transient electron dynamics in
atomic~\cite{goulielmakis_real-time_2010},
molecular~\cite{lepine_molecular_2013} and
condensed~\cite{cavalieri_attosecond_2007} matter. The attosecond
streaking technique in particular captures the 
\emph{time-resolved} photoelectron spectra and thus allows for tracing
the pathway of plasmon-accompanied photoemission in the time domain
\cite{neppl_direct_2015,lemell_real-time_2015,lucchini_light-matter_2015}.

Generally, photoemission is an involved
process~\cite{hufner_photoelectron_2003} in which several factors are
important: the density of states of the unperturbed system, the
electron scattering following photoabsorption and the
formation of electron scattering states which are subsequently
observed in the detector. The corresponding three stages are known as
the classical model of photoemission due to Berglund and
Spicer~\cite{berglund_photoemission_1964}. The last stage is
complicated by the presence of long-range Coulomb interaction
between the emitted particle and the target. Fortunately, in many
cases calculations of the scattering states can be decoupled from the
treatment of the many-body effects, which is the main topic of this
work.

A deep core hole is created due to the interaction with an XUV
photon. The liberated electron interacts with both, the particle-hole
excitations in the conduction band and may also excite collective
charge density fluctuations (plasmons). The separation between these
scattering mechanisms is only possible in the long wave-length limit
where particle-hole excitations shape the threshold profile. The
plasmons, by acting as massive bosonic particles, reshape the satellites
features in the spectrum~\cite{langreth_singularities_1970}. The
latter effect, which is inherently nonequilibrium and known as
\emph{extrinsic} losses, should be distinguished from the
\emph{intrinsic} losses manifested as, e.\,g. plasmonic satellites
(PSs) in the equilibrium spectral function.  The occurrence of quantum
interference between these two channels is essential for obtaining
accurate photoemission spectra in the vicinity of PSs
\cite{campbell_interference_2002}.  A microscopic theory accounting
for intrinsic and extrinsic losses is a challenge even in standard
steady-state photoemission theory
\cite{guzzo_valence_2011,klevak_charge_2014,lischner_physical_2013,
  pavlyukh_single-or_2015}, while a time-dependent description is
still lacking.

In this work we focus on the time-dependent aspects of photoemission
for electronic systems where the interaction is solely mediated by the
bosonic excitations. Typical examples are processes involving
electron-phonon or electron-photon interactions. Also in pure
electronic systems the interaction can often be written in this form:
e.\,g. for deep core photoemission the photoelectron at high energies
can be treated as distinguishable particle interacting with the
density fluctuation of the system~\cite{hedin_transition_1998}. At
metallic densities fluctuations are dominated by plasmonic
excitations. This gives rise to the s-model originally proposed by
Lundqvist \cite{lundqvist_characteristic_1969} and solved by
Langreth~\cite{langreth_singularities_1970}. Keeping in mind the
distinguishability aspect of such a reduction the model can also be
applied to more general scenarios such as homogeneous electron gas at
metallic densities~\cite{minnhagen_aspects_1975} or solids treated in
the plasmon-pole approximation~\cite{aryasetiawan_gw_1998}. A sequence
of plasmonic satellites accompanying the main quasi-particle (QP) peak
is a generic feature of the density of states of the electron-boson
Hamiltonian~\cite{cini_exactly_1988}.

A powerful method to deal with time-dependent process in many body
systems is the nonequilibrium Green function (NEGF) approach. It
provides a link to standard many-body perturbation theory, allowing so
for systematic approximation schemes, and also to classical kinetics
\cite{lipavsky_generalized_1986}.
On important application of the NEGF formalism is the prediction and
interpretation of time-resolved angular-resolved photoemission spectra
(tr-ARPES) \cite{freericks_theoretical_2009}\,---\,a technique that is
used in recent experiments on ultrafast dynamics of electronic
\cite{sohrt_how_2014} or phononic \cite{wall_ultrafast_2012} band
structures of correlated materials.

The method relies on solving the equations of motion (EOM) for the
Green's functions on the Keldysh
time-contour~\cite{dahlen_solving_2007,myohanen_kadanoff-baym_2009,balzer_nonequilibrium_2012}
---\,the Kadanoff-Baym equations (KBEs)\,---\,with a proper choice of
self-energy~\cite{von_friesen_successes_2009,puig_von_friesen_kadanoff-baym_2010,verdozzi_open_2011},
which in turn determines the form of collision integrals.  This work
is devoted to the extension of this formalism to coupled
electron-boson Hamiltonians (Sec.~\ref{subsec:modelham}) and
formulation of bosonic EOM as a second order equation for massive
particles (Sec.~\ref{subsec:eom}). The formalism is kept general and
is thus applicable to related problems such as pseudo-particles
\cite{white_inelastic_2012}, electron-phonon
\cite{mahan_many-particle_2000,galperin_line_2004,
  dash_nonequilibrium_2010,sukharev_transport_2010,ness_nonequilibrium_2011},
electron-vibron \cite{schuler_local_2013,bostrom_time-resolved_2015}
or electron-photon \cite{pellegrini_optimized_2015}, or plasmonic
nanojunctions \cite{white_collective_2012,kaasbjerg_theory_2015}.  In
this study we go beyond the frozen boson scheme as often employed for
electron-phonon coupling
\cite{sentef_examining_2013,eckstein_photoinduced_2013} and treat
density oscillations in the system quantum mechanically.  Our
time-dependent numerical approach (Sec.~\ref{sec:numerics}) allows to
disentangle intrinsic from extrinsic losses in photoemission in a
natural way and complement the steady regime studies that have been
performed previously~\cite{lischner_physical_2013,guzzo_multiple_2014,
  vinson_bethe-salpeter_2011,kas_real-time_2015,klevak_charge_2014,kas_many-pole_2007,almbladh_comments_1978,
  almbladh_importance_1986,inglesfield_plasmon_1983,penn_role_1978,penn_theory_1977,almbladh_theory_1985,
  fujikawa_new_2009,chang_deep-hole_1973,fujikawa_many-body_2002,hedin_external_2001,uimonen_ultra-nonlocality_2014}.

We apply the theory to the time-resolved photoemission from the
magnesium 2p core state and discuss the influence of intrinsic and
extrinsic electron-plasmon couplings (Sec.~\ref{sec:appl}). Atomic units
are used unless stated otherwise.

\section{Theory \label{sec:theory}}

Our goal is the description of a system of electrons interacting with
bosonic QPs that can be emitted or absorbed and thus
mediate an effective electron-electron interaction. As a consequence,
the boson propagators must have spectral features that are quite
different from (non-relativistic) electrons: instead of one QP peak at
energy $\mathcal{E}$, a boson mode with frequency $\Omega$ is
represented by two peaks at $\pm \Omega$ in the spectral function
$\hat{B}(\omega)$, corresponding to emission or absorption of the QP,
respectively. More generally, this is reflected by the anti-symmetry
of the boson spectral function
$\hat{B}(\omega)=-\hat{B}^\mathrm{T}(-\omega)$, which make it
different from that of real bosonic particles such as atoms with
integer nuclear spin.

\subsection{Generic Hamiltonian \label{subsec:modelham}}

Let us consider a system characterized by a set of electronic
single-particle (SP) states with energies $\{\mathcal{E}_i\}$ and
possessing a number of boson modes with corresponding frequencies
$\{\Omega_\nu\}$. The respective annihilation operators of the
electrons (bosons) are denoted by $\hat{c}_i$
($\hat{a}_\nu$).

For the electrons we have the non-interacting Hamiltonian
\begin{equation}
  \hat{H}_\mathrm{el} = \sum_i \mathcal{E}_i \, \hat{c}^\dagger_i
  \hat{c}_i \ ,
\end{equation}
while
\begin{equation}
  \hat{H}_\mathrm{bos} = \sum_\nu \Omega_\nu\, \hat{a}^\dagger_\nu \hat{a}_\nu
  = \frac12 \sum_\nu \Omega_\nu 
  \left(\hat{P}^2_\nu +\hat{Q}^2_\nu\right)  
\end{equation} 
represents the boson Hamiltonian. Instead of working with the bosonic
creation or annihilation  operators
the coordinate-momentum representation,
\begin{equation}
  \label{eq:bosonPQ}
  \hat{Q}_\nu = \frac{1}{\sqrt{2}} \left(\hat{a}_\nu + \hat{a}^\dagger_\nu\right) 
  \ , \quad
  \hat{P}_\nu = \frac{1}{\sqrt{2}\iu} \left(\hat{a}_\nu - \hat{a}^\dagger_\nu\right)  \ ,
\end{equation}
is preferred here. Note that electrons and bosons (besides their
coupling) are considered as non-interacting here for the sake of a
clear presentation. However, additional correlation effects for both
subsystems can, in principle, be included without conceptional
obstacles.

The electron-boson interaction is taken as
\begin{equation}
  \label{eq:Hel-bos}
  \hat{H}_\mathrm{el-bos} = \sum_\nu\sum_{ij}
  \Gamma^\nu_{ij}\hat{c}^\dagger_i\hat{c}_j 
  \hat{Q}_\nu \ .
\end{equation}
A coupling were the order of the fermionic operators is interchanged (e.\,g. $\hat{c}_i
\hat{c}^\dagger_j$) can be treated along the same lines by employing the anti-commutator
relation $\hat{c}_i \hat{c}^\dagger_j = \delta_{ij}-\hat{c}^\dagger_j \hat{c}_i$. The
remaining term arising due to the Kronecker delta, $\sum_{\nu,i}\Gamma^\nu_{ii}
\hat{Q}_\nu$ can be removed by shifting bosonic coordinates.

Furthermore we account for environmental effects such as particle
exchange and line broadening by including additional baths.  In
analogy to above, we define the environment SP states by the energies
$\{\epsilon_k \}$ whereas the boson bath is characterized by the
frequencies $\{\omega_\alpha\}$.
\begin{equation}
  \hat{H}^\mathcal{B}_\mathrm{el} = \sum_k \epsilon_k\, \hat{d}^\dagger_k\hat{d}_k \ ,
  \quad
  \hat{H}^\mathcal{B}_\mathrm{bos} = \frac12 \sum_\alpha \omega_\alpha 
  \left(\hat{p}^2_\alpha + \hat{q}^2_\alpha\right) \ .
\end{equation}
The boson bath operators $\hat{p}_\alpha$, $\hat{q}_\alpha$ are
defined analogous to eq.~\eqref{eq:bosonPQ}, while $\hat{d}_k$ denotes
the annihilation operators with respect to the electron bath. The
coupling of the electron-boson system to the environmental degrees of
freedom is described by the embedding Hamiltonians
\begin{equation}
  \hat{H}_\mathrm{el-em} = \sum_{i,k} \left(V_{ik}
    \hat{c}^\dagger_i \hat{d}_k +\mathrm{h. c.} \right) 
\end{equation}
and
\begin{equation}
  \hat{H}_\mathrm{bos-em} = \sum_{\alpha,\nu} 
  \gamma_{\alpha,\nu} \hat{Q}_\nu \hat{q}_\alpha \ .
\end{equation}
The total static Hamiltonian thus reads
\begin{equation}
  \label{eq:H0}
  \hat{H}_0 = \hat{H}_\mathrm{el}+\hat{H}_\mathrm{bos}+\hat{H}_\mathrm{el-bos}
  + \hat{H}^\mathcal{B}_\mathrm{el}+\hat{H}^\mathcal{B}_\mathrm{bos}
  + \hat{H}_\mathrm{el-em} + \hat{H}_\mathrm{bos-em} \ .
\end{equation}
For later convenience we also introduce 
\begin{equation}
  \label{eq:H0ad}
  \hat{H}^\prime_0(t) = \hat{H}_\mathrm{el}+\hat{H}_\mathrm{bos} 
  + s(t) \hat{H}_\mathrm{int} \ ,
\end{equation}
where $\hat{H}_\mathrm{int}$ comprises all the interacting
contributions from eq.~\eqref{eq:H0}. The modified Hamiltonian
\eqref{eq:H0ad} allows, by choosing a suitable functional form for
the scaling factor $s(t)$, to "switch on" the interaction adiabatically
in order to obtain fully correlated eigenstates of $\hat{H}_0$, while
$s\equiv 1$ retrieves the static case.

To account for light-matter interaction we introduce 
\begin{equation}
  \label{eq:Hel-L}
  \hat{H}_\mathrm{el-L}(t) = \sum_{ij} F_{ij}(t) \hat{c}^\dagger_i \hat{c}_j  + \mathrm{h. c. } \ ,
\end{equation}
where $F_{ij}(t)$ comprise the transition matrix elements and the time-dependent
field. There is no direct coupling of light to bosonic excitations in the minimal coupling
scheme. The total time-dependent Hamiltonian is then given by
\begin{equation}
  \label{eq:Htot}
  \hat{H}(t) = \hat{H}^\prime_0(t) + \hat{H}_\mathrm{el-L}(t) \ .
\end{equation}

\subsection{Equations of motion \label{subsec:eom}}
To treat photoemission for the system described by the
Hamiltonian~\eqref{eq:Htot} we proceed in a standard way by
considering the one-particle fermionic and bosonic Green's
functions~\cite{pal_conserving_2009,pavlyukh_single-or_2015}.
Transient optical absorption requires the use of more complicated
two-particle Green's
functions~\cite{pal_optical_2011,moskalenko_attosecond_2012,perfetto_non-equilibrium_2015}
and is outside of the scope of this manuscript.

\begin{figure}[t]
  \centering
  \includegraphics[width=0.6\columnwidth]{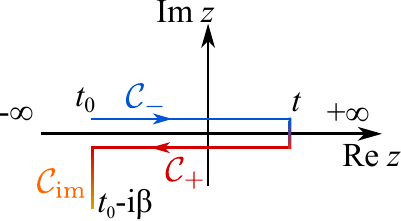}
  \caption{The general contour $\mathcal{C}$ consisting of the forward
    branch $\mathcal{C}_-$ on the real axis, the backward branch $\mathcal{C}_+$
    and the imaginary branch $\mathcal{C}_\mathrm{im}$. The arrows indicate
    the direction of the contour-ordering. $\beta$ denotes the inverse
    temperature. \label{fig1}}
\end{figure}

Thus, let us introduce the electron GF
\begin{equation}
  G_{ij}(z_1,z_2) = -\iu \langle \mathcal{T} \hat{c}_i(z_1) \hat{c}^\dagger_j(z_2) \rangle \ .
\end{equation}
Here, $z_1$ and $z_2$ are time arguments on the general contour
$\mathcal{C}$\cite{stefanucci_nonequilibrium_2013} (sketched in
fig.~\ref{fig1}), while $\mathcal{T}$ represents the corresponding
contour-ordering operator. All operators are represented in a contour
Heisenberg picture. The average $\langle \dots \rangle$ refers to an
initial ensemble of eigenstates of the Hamiltonian
$\hat{H}^\mathrm{M}$. Typical choices here are (i)
$\hat{H}^\mathrm{M}=\hat{H}_0 - \mu \hat{n}_\mathrm{el}$
($\hat{n}_\mathrm{el}$ is the electron number operator), or (ii)
$\hat{H}^\mathrm{M}=\hat{H}_\mathrm{el} + \hat{H}_\mathrm{bos} - \mu
\hat{n}_\mathrm{el}$. Case (i) prepares the system in an ensemble with
initial correlation, whereas in (ii) the non-interacting and thus
known basis is used as reference. Adiabatic switching can then be
employed to obtain a correlated state by turning on the interaction
along the real time-axis \cite{balzer_nonequilibrium_2012}. Note that
the chemical potential for the bosons is assumed to be zero as, in
principle, an infinite number of them can be created.

Next, we define the boson GF\,---\,the coordinate-coordinate correlator\,---\,according to
\begin{equation}
  \label{eq:plasmongf1}
  \begin{split}
    D_{\mu\nu}(z_1,z_2) &= -\iu \big[\langle \mathcal{T}\hat{Q}_\mu(z_1)\hat{Q}_\nu(z_2)\rangle
    -\langle\hat{Q}_\mu(z_1)\rangle\langle\hat{Q}_\nu(z_2)\rangle
    \big] \\
    &= -\iu \langle
    \Delta\hat{Q}_\mu(z_1)\Delta\hat{Q}_\nu(z_2)\rangle 
  \end{split}
\end{equation}
with the fluctuation operator $\Delta\hat{Q}_\nu(z) =\hat{Q}_\nu(z)-\langle \hat{Q}_\nu(z)
\rangle$. Likewise the momentum-coordinate
\begin{equation}
  D^{PQ}_{\mu\nu}(z_1,z_2) = -\iu \big[\langle \mathcal{T} \hat{P}_\mu(z_1) \hat Q_\nu(z_2) \rangle
  - \langle \hat P_\mu(z_1) \rangle \langle \hat Q_\nu(z_2) \rangle \big]\ .
\end{equation}
and momentum-momentum correlators 
\begin{equation}
  D^{PP}_{\mu\nu}(z_1,z_2) = -\iu \big[\langle \mathcal{T} \hat{P}_\mu(z_1) \hat P_\nu(z_2) \rangle
  - \langle \hat P_\mu(z_1) \rangle \langle \hat P_\nu(z_2) \rangle \big]\ .
\end{equation}
can be defined. We will demonstrate below that they are not required for the propagation
of $D_{\mu\nu}(z_1,z_2)$, but are necessary if one is interested in the observables such
as bosonic occupation number.

In order to elucidate the features of the respective self-energies
related to the explicit time dependence, we have re-derived the EOM
using the source-field method \cite{strinati_application_1988}. The
full derivation is presented in appendix~\ref{app:eom}. Here we
recapitulate the key points.

The electron GF (represented as matrix) obeys, as usual,
\begin{equation}
  \label{eq:eomgfel1}
  \left(\iu \frac{\partial}{\partial z_1}\mathbf{I}-\mathbf{h}^\mathrm{MF}(z_1)\right)
  \mathbf{G}(z_1,z_2)
  =\delta(z_1,z_2) +\int_\mathcal{C}\dd z_3\,\mathbf{\Sigma}(z_1,z_3)
  \mathbf{G}(z_3,z_2) \ .
\end{equation}
The self-energy, comprising many-body effects due to the
electron-boson interaction and the coupling to the environment,
appears as a mean-field (MF) contribution incorporated into the MF
Hamiltonian $\mathbf{h}^\mathrm{MF}$,
\begin{align}
  \label{eq:hmf1}
  h^\mathrm{MF}_{ik}(z) = \mathcal{E}_i \delta_{ik} + F_{ik}(z) + s(z)\sum_\nu \Gamma^\nu_{ik} 
  \langle \hat{Q}_\nu(z) \rangle
\end{align}
and as the time non-local correlation self-energy.

The EOM for the boson propagator $D_{\mu\nu}(z_1,z_2)$ can be derived (details in
appendix~\ref{app:eom}) from the Heisenberg EOM for position and momentum operators:
\begin{align}
  \label{eq:HeisEOMQ}
  \frac{\dd}{\dd z} \hat{Q}_\nu(z) &= \Omega_\nu \hat{P}_\nu(z)
  \ , \\
    \label{eq:HeisEOMP} \frac{\dd}{\dd z} \hat{P}_\nu(z) &= -\Omega_\nu \hat{Q}_\nu
                                      (z) - s(z)\sum_{ij} \Gamma^\nu_{ij}
                                      \hat{c}^\dagger_i(z) \hat{c}_j(z) \\
  &\quad-s(z)\sum_\alpha \gamma_{\alpha,\nu} \hat{q}_\alpha(z) \nonumber \ .
\end{align}
They show that the first order equation for $D_{\mu\nu}(z_1,z_2)$ involves
momentum-position correlators $D_{\mu\nu}^{PQ}$ and can only be closed as a second-order
equation.  The notion of the (boson) self-energy $\Pi_{\mu\nu}(z_1,z_2)$, in the same
spirit as for electronic GFs, results from closing the EOM. Gathering environmental
and polarization effects into $\Pi_{\mu\nu}(z_1,z_2)$, the contour EOM
for the boson GF reads
\begin{align}
  \label{eq:EOMDnu}
  -\frac{1}{\Omega_\nu} \left(\frac{\partial^2}{\partial z^2_1} +
    \Omega^2_\nu\right) D_{\mu\nu}(z_1,z_2) &=\delta_{\mu\nu}\delta(z_1,z_2)\nonumber \\
    &\hspace{-5em}+\sum_\xi\int_\mathcal{C}\!\dd z_3\,\Pi_{\mu\xi}(z_1,z_3)  D_{\xi\nu}(z_3,z_2) \ .
\end{align}
In contrast to the electron case \eqref{eq:eomgfel1}, the boson
propagators are subject to a second-order EOM. 

For examining the
spectral properties we define the different Keldysh components
depending on which branch of the contour (fig.~\ref{fig1}) the
arguments $(z_1,z_2)$ are located (we adopt the conventions from
ref.~\onlinecite{stefanucci_nonequilibrium_2013}). E.\,g. the
greater/lesser boson GF $D^\gtrless_{\mu\nu}(t_1,t_2)$ corresponds to
$D_{\mu\nu}(z_1,z_2)$ with $z_1=t_1 \in \mathcal{C}_\pm$ and $z_2=t_2
\in \mathcal{C}_\mp$. In equilibrium, $D^\gtrless_{\mu\nu}(t_1,t_2)$
depends on $t_1-t_2$ only, allowing to perform the Fourier
transformation $D^\gtrless_{\mu\nu}(\omega)=\int^\infty_{-\infty}\dd
t\, e^{\iu \omega t}D^\gtrless_{\mu\nu}(t)$.  For instance, the
resulting spectral function for the non-interacting case $\Pi_{\mu\nu}
= 0$ reads $b_{\mu\nu}(\omega) = \pi\delta_{\mu\nu}
\left[\delta(\omega-\Omega_\nu)- \delta(\omega+\Omega_\nu)\right]$.
The appearance of the two peaks is, as mentioned above, a consequence
of the second-order EOM eq.~\eqref{eq:EOMDnu}. Further properties of
the boson propagators are summarized in appendix~\ref{app:basic}.

As detailed in appendix~\ref{app:eom}, the expression for the electron self-energy due to
the electron-boson interaction is given by
\begin{align}
  \Sigma^\mathrm{el-bos}_{ij}(z_1,z_2)&=\iu\, s(z_1)\sum_{\mu\nu}
  \sum_{nk}\sum_{ab}\Gamma^\mu_{ik} \Gamma^\nu_{ab}\int_\mathcal{C}\dd (z_3 z_5)
  G_{kn}(z_1,z_3)\nonumber\\ &\quad\quad \times \Lambda_{njab}(z_3,z_2;z_5)
  D_{\mu\nu}(z_5,z^+_1) s(z_5) \ ,
\end{align}
where $\Lambda_{njab}$ denotes the three-point vertex function obeying the standard
Bethe-Salpeter equation (BSE) with the four-point kernel $K_{abcd}(z_1,z_2;z_3,z_4)=\delta
\Sigma^\mathrm{el-bos}(z_1,z_2)/\delta G_{cd}(z_3,z_4)$ obtained from the functional
derivative of the self-energy with respect to the electron GF (details in
appendix~\ref{app:eom}). The boson self-energy in turn is determined by the electron
(irreducible) polarization,
\begin{equation}
  \label{eq:polgglambda}
  \begin{split}
    P_{abcd}(z_1,z_2) &= -\iu \sum_{pq} \int_\mathcal{C}\dd (z_3 z_4)G_{ap}(z_1,z_3) G_{qb}(z_4,z^+_1)\\
    &\quad \times \Lambda_{pqcd}(z_3,z_4;z_2) \ ,
  \end{split}
\end{equation}
by 
\begin{equation}
  \Pi^\mathrm{p}_{\mu\nu}(z_1,z_2)=s(z_1)s(z_2)\sum_{abcd}\Gamma^\mu_{ba}P_{abcd}(z_1,z_2)\Gamma^\nu_{cd} \ .
\end{equation}
The simplest possible conserving approximation
\cite{baym_conservation_1961} emerges from invoking the zeroth-order
approximation to the vertex function, that is
\begin{equation}
  \label{eq:vertex0}
  \Lambda_{abcd}(z_1,z_2;z_3) = \delta_{ac}\delta_{bd}\delta(z_1,z_2)\delta(z_1,z_3) \ .
\end{equation}
\begin{figure}[t]
  \centering
  \includegraphics[width=\columnwidth]{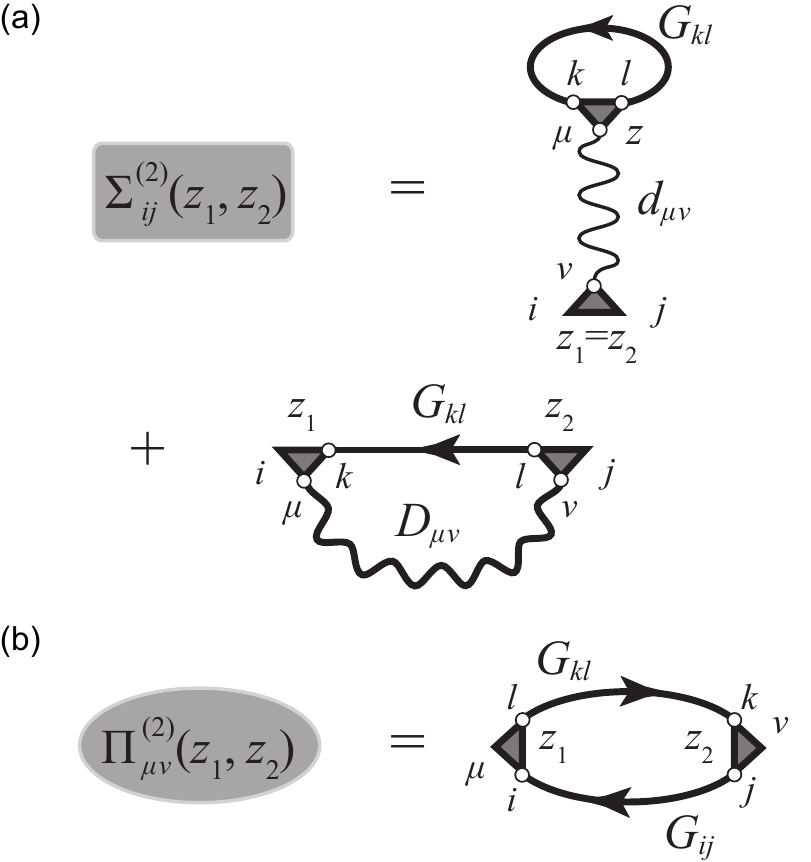}
  \caption{Approximations for the fermionic (a) and bosonic (b) self-energy operators
    employed in this work. All self-energies are of the second order in
    $\mathbf{\Gamma}^\mu$ and are expressed in terms of full electronic $\mathbf{G}$ and
    bare $d_{\mu\nu}$ and full $D_{\mu\nu}$ boson propagators. The first term in the
    fermionic self-energy is local in time and therefore is included (see
    Eq.~\eqref{eq:hmf1}) in mean-field Hamiltonian $\mathbf{h}^\mathrm{MF}$.
    \label{fig:diag}}
\end{figure}
Analogously to Hedin's equations for electronic systems
(fig.~\ref{fig:diag}), we designate the resulting second-order (in
$\mathbf{\Gamma}$) approximations to both the electron and the boson
self-energy as $GW$ approximation:
\begin{subequations}
  \label{eq:GWA1}
  \begin{equation}
    \mathbf{\Sigma}^{(2)}(z_1,z_2) = \iu\, s(z_1)s(z_2)\sum_{\mu\nu} \mathbf{\Gamma}^\mu\mathbf{G}(z_1,z_2)
    \mathbf{\Gamma}^\nu D_{\mu\nu}(z_1,z_2) \ ,
  \end{equation}
  \begin{equation}
   \Pi^{(2)}_{\mu\nu}(z_1,z_2) = 
    -\iu \, s(z_1) s(z_2)\mathrm{Tr}\left[\mathbf{\Gamma}^\mu\mathbf{G}(z_1,z_2)
      \mathbf{\Gamma}^\nu\mathbf{G}(z_2,z_1)\right] \ .
  \end{equation}
\end{subequations}
The contribution to the respective self-energies arising from the
environmental coupling (embedding self-energies) are expressed in the
standard way in terms of the bath propagators for electrons:
\begin{equation}
  \label{eq:sgmem1}
  \Sigma^\mathrm{em}_{ij}(z_1,z_2)= s(z_1) s(z_2) \sum_{k} V_{ik}V^*_{kj}
  g^\mathcal{B}_{k}(z_1,z_2) \ ,
\end{equation}
and for bosons
\begin{equation}
  \label{eq:piem1}
  \Pi^\mathrm{em}_{\nu}(z_1,z_2)= s(z_1) s(z_2)\sum_{\alpha}
  \left|\gamma_{\alpha,\nu}\right|^2
  d^\mathcal{B}_\alpha(z_1,z_2) \ .
\end{equation}
Here, $g^\mathcal{B}_k$ and $d^\mathcal{B}_\alpha$ are the bare GFs of
the respective baths.  It should be noted that here the boson
embedding self-energy is labeled by a single mode index $\nu$. In
general, non-diagonal terms can occur due to indirect coupling via the
bath. Thus, eq.~\eqref{eq:piem1} relies on the assumption that such
effects can be neglected.  As usual, the full self-energy is obtained
by summing the system and the bath contributions, that is
$\mathbf{\Sigma} =\mathbf{\Sigma}^\mathrm{el-bos} +
\mathbf{\Sigma}^\mathrm{em}$ and $\Pi_{\mu\nu} =
\Pi^\mathrm{p}_{\mu\nu}+\Pi^\mathrm{em}_{\mu\nu}$, respectively.

Eq.~\eqref{eq:EOMDnu} for the coordinate-coordinate correlator
$D_{\mu\nu}(z_1,z_2)$ is not sufficient to fully describe the boson
dynamics, as the MF Hamiltonian \eqref{eq:hmf1} explicitly depends on $\langle
\hat{Q}_\nu(z)\rangle$ (this quantity can not be inferred from
$D_{\mu\nu}(z_1,z_2)$). An additional EOM is therefore required and
can be derived from eq.~\eqref{eq:HeisEOMQ} and
\eqref{eq:HeisEOMP}. Eliminating the bath amplitudes
$\hat{q}_\alpha(z)$ by the standard embedding technique, one obtains
\begin{align}
  \label{eq:EOMQ}
   -\frac{1}{\Omega_\nu}\left(\frac{\partial^2}{\partial z^2} +
    \Omega^2_\nu\right) \langle \hat{Q}_\nu(z) \rangle &= -\iu \, 
  \mathrm{Tr}\left[\mathbf{\Gamma}^\nu \mathbf{G}(z,z^+)\right] \nonumber \\
  &\quad +\int_\mathcal{C}\dd \bar{z}\, \Pi^\mathrm{em}_\nu(z,\bar{z})
  \langle \hat{Q}_\nu(\bar{z}) \rangle
\end{align}
The EOM~\eqref{eq:eomgfel1}, \eqref{eq:EOMDnu}, \eqref{eq:EOMQ} for
quantities on the contour is to be solved together with
Kubo-Martin-Schwinger (KMS) boundary
conditions~\cite{kadanoff_quantum_1994}. For eq.~\eqref{eq:EOMQ} this
implies that the solution is separated into a boundary-value problem
for $z=-\iu \tau \in \mathcal{C}_\mathrm{im}$, as $\langle
\hat{Q}_\nu(0) \rangle = \langle \hat{Q}_\nu(-\iu \beta) \rangle$,
whereas for $z \in \mathcal{C}_\pm$ eq.~\eqref{eq:EOMQ} represents an
initial-value problem.

In absence of environmental coupling (i.\,e. $\Pi^\mathrm{em}_\nu=0$),
eq.~\eqref{eq:EOMQ} can be solved in terms of the non-interacting
boson propagators $d_\nu(z_1,z_2)$, yielding
\begin{align}
  \label{eq:EOMQslv}
  \langle \hat{Q}_\nu(z) \rangle = -\iu\int_\mathcal{C}\dd\bar{z}\, d_\nu(z,\bar{z}) 
  \mathrm{Tr}\left[\mathbf{\Gamma}^\nu \mathbf{G}(\bar z,\bar{z}^+)\right] \ .
\end{align}
Substituting eq.~\eqref{eq:EOMQslv} back into the MF Hamiltonian
\eqref{eq:hmf1} we obtain the first diagram depicted in
fig.~\ref{fig:diag}(a). Thus, the first-order (in
$\mathbf{\Gamma}^\nu$) MF expression has been
transformed to a (formally) second-order self-energy, which is often
referred to as Hartree term
\cite{viljas_electron-vibration_2005,dash_nonequilibrium_2010}. We
stress that this transition is not possible in presence of a boson
bath ($\Pi^\mathrm{em}_\nu\neq 0$). The MF part will hence be kept in
the more general form eq.~\eqref{eq:hmf1}.  This is analogous to
ref.~\onlinecite{sakkinen_many-body_2015}.

Furthermore, propagating the boson amplitude $\langle \hat{Q}_\nu(z)
\rangle$ is necessary for computing the boson occupation number
$N_\nu$:
\begin{align}
  \label{eq:occbos} 
  N_\nu(z) &= \langle \hat{a}^\dagger_\nu(z) \hat{a}_\nu(z) \rangle = 
  \frac12\left[\langle \hat{P}_\nu(z)^2 \rangle +\langle \hat{Q}_\nu(z)^2 
    \rangle -1 \right] \nonumber \\
  &= \frac{\iu}{2}\left[D_{\nu\nu}(z,z^+)+D^{PP}_{\nu\nu}(z,z^+)\right] \\
  &\quad +\frac12\left[\langle \hat{P}_\nu(z) \rangle^2+ 
    \langle \hat{Q}_\nu(z) \rangle^2 -1 \right] \nonumber \ .
\end{align}

\section{Numerical implementation \label{sec:numerics}}

In this section we revisit the formulation of the KBE from the contour
EOM. Since the general solution strategy in case of electron GFs is
quite established
\cite{dahlen_solving_2007,myohanen_kadanoff-baym_2009,balzer_nonequilibrium_2012},
we keep the discussion brief and rather focus on the modifications to
be made for calculating the bosonic time-evolution.

Together with the corresponding adjoint EOM, eq.~\eqref{eq:eomgfel1}
and \eqref{eq:EOMDnu} represent the KBEs for the coupled
electron-boson system that needs to be solved along with
eq.~\eqref{eq:EOMQ}. For a numerical approach, the general complex
contour arguments are mapped onto observable times by splitting the
general GFs into their respective Keldysh components. Let us introduce
the convolution operations
\begin{align}
  [f\cdot g](t,t') &\equiv \int^\infty_{t_0}\!\dd \bar{t}\, f(t,\bar t) g(\bar t,t') \ ,\\
  [f\star g](t,t')&\equiv -\iu \int^\beta_{0}\!\dd \bar{\tau}\, 
  f(t,\bar{\tau}) g(\bar{\tau},t') \ .
\end{align}
Applying the Langreth rules \cite{stefanucci_nonequilibrium_2013}, the
KBEs for the greater/lesser electron GF becomes
\begin{subequations}
  \label{eq:KBEel}
  \begin{equation}
    \label{eq:KBEel1}
    \iu \frac{\partial}{\partial t_1} \mathbf{G}^\gtrless(t_1,t_2)
    =\mathbf{h}^\mathrm{MF}(t_1) \mathbf{G}^\gtrless(t_1,t_2)+ \mathbf{X}^\gtrless_L(t_1,t_2) \ ,
  \end{equation}
  \begin{equation}
    \label{eq:KBEel2}
     -\iu \frac{\partial}{\partial t_2} \mathbf{G}^\gtrless(t_1,t_2)
    = \mathbf{G}^\gtrless(t_1,t_2)\mathbf{h}^\mathrm{MF}(t_2)+ \mathbf{X}^\gtrless_R(t_1,t_2) \ ,
  \end{equation}
  \begin{equation}
    \label{eq:KBEel3}
    \iu \frac{\partial}{\partial t} \mathbf{G}^\rceil(t,\tau)
    =\mathbf{h}^\mathrm{MF}(t) \mathbf{G}^\rceil(t,\tau)+ \mathbf{X}^\rceil_L(t,\tau) \ ,
  \end{equation}
\end{subequations}
with the standard collision integrals
\begin{subequations}
  \label{eq:collx}
  \begin{equation}
    \mathbf{X}^\gtrless_L(t_1,t_2)=\left[
      \mathbf{\Sigma}^\mathrm{R}\cdot\mathbf{G}^\gtrless
      +\mathbf{\Sigma}^\mathrm{\gtrless}\cdot\mathbf{G}^\mathrm{A}
      +\mathbf{\Sigma}^\rceil\star\mathbf{G}^\lceil
    \right](t_1,t_2) \ ,
  \end{equation}
   \begin{equation}
     \mathbf{X}^\gtrless_R(t_1,t_2)=\left[
      \mathbf{G}^\mathrm{R}\cdot\mathbf{\Sigma}^\gtrless
      +\mathbf{G}^\mathrm{\gtrless}\cdot\mathbf{\Sigma}^\mathrm{A}
      +\mathbf{G}^\rceil\star\mathbf{\Sigma}^\lceil
    \right](t_1,t_2) \ ,
  \end{equation}
  \begin{equation}
    \mathbf{X}^\rceil_L(t,\tau)=\left[
      \mathbf{\Sigma}^\mathrm{R}\cdot \mathbf{G}^\rceil
      +\mathbf{\Sigma}^\rceil\star \mathbf{G}^\mathrm{M}
      \right](t,\tau) \ .
  \end{equation}
\end{subequations}
Similarly to eq.~\eqref{eq:KBEel} one finds the KBEs for the boson
propagators,
\begin{subequations}
  \label{eq:KBEbos}
  \begin{equation}
    \label{eq:KBEbos1}
    -\frac{1}{\Omega_\mu}\left(\frac{\partial^2}{\partial t^2_1}
      +\Omega^2_\mu\right) D^\gtrless_{\mu\nu}(t_1,t_2) = Y^\gtrless_{L,\mu\nu}(t_1,t_2) \ ,
  \end{equation}
   \begin{equation}
    \label{eq:KBEbos2}
    -\frac{1}{\Omega_\nu}\left(\frac{\partial^2}{\partial t^2_2}
      +\Omega^2_\nu\right) D^\gtrless_{\mu\nu}(t_1,t_2) = Y^\gtrless_{R,\mu\nu}(t_1,t_2) \ ,
  \end{equation}
  \begin{equation}
    \label{eq:KBEbos3}
    -\frac{1}{\Omega_\mu}\left(\frac{\partial^2}{\partial t^2}
      +\Omega^2_\mu\right) D^\rceil_{\mu\nu}(t,\tau) = Y^\rceil_{L,\mu\nu}(t,\tau) \ ,
  \end{equation}
\end{subequations}
where the collision integrals are obtained by applying the Langreth
rules analogously as in eq.~\eqref{eq:collx}. The symmetry properties
of the GFs (and of the respective self-energies) lead to similar
relations for the collision integrals:
\begin{align*}
  \mathbf{X}^\gtrless_L(t_1,t_2) = - \left[\mathbf{X}^\gtrless_R(t_2,t_1)\right]^\dagger \ ,
  \quad Y^\gtrless_{L,\mu\nu}(t_1,t_2) = -\left[Y^\gtrless_{R,\nu\mu}(t_2,t_1)\right]^* \ .
\end{align*}

Let us now assume that the Matsubara GF for both electrons
($\mathbf{G}^\mathrm{M}(\tau)$) and bosons ($D^\mathrm{M}_{\mu\nu}(\tau)$)
has been determined by solving the respective Matsubara Dyson
equation. Note that the solution has to be carried out
self-consistently with the EOM for the boson amplitude
(eq.~\eqref{eq:EOMQ}). From the KMS conditions one finds
\begin{align}
  \label{eq:EOMQim}
  \frac{1}{\Omega_\nu}\left(\frac{\dd^2}{\dd \tau^2}-\Omega^2_\nu\right)
   Q^\mathrm{M}_\nu(\tau) &= -\iu\,\mathrm{Tr}\left[
  \mathbf{\Gamma}^\nu \mathbf{G}^\mathrm{M}(0)\right] \\  
&\quad + \left[\Pi^\mathrm{em,M}_\nu \star 
 Q^\mathrm{M}_\nu \right](\tau)  \nonumber \ .
\end{align}
Here, $Q^\mathrm{M}_\nu(\tau) = \langle \hat Q_\nu(t_0-\iu
\tau)\rangle $.  Solving the imaginary track $\mathcal{C}_\mathrm{im}$
is straightforward if one discards initial correlations, as done in
the adiabatic switching method. Once the GFs at $t_0$ have been
initialized from the Matsubara components, eq.~\eqref{eq:KBEel} and
\eqref{eq:KBEbos} can be propagated for real times together with the
boson amplitude
\begin{align}
  \label{eq:EOMQre}
  -\frac{1}{\Omega_\nu}\left(\frac{\dd^2}{\dd t^2}+\Omega^2_\nu\right)
  \langle \hat Q_\nu(t)\rangle &= -\iu\,\mathrm{Tr}\left[
  \mathbf{\Gamma}^\nu \mathbf{G}^<(t,t)\right] \nonumber \\  
&\quad +\left[\Pi^\mathrm{em,R}_\nu \cdot
\langle \hat Q_\nu\rangle \right](t) \ .
\end{align}

\begin{figure}[t]
  \centering
  \includegraphics[width=\columnwidth]{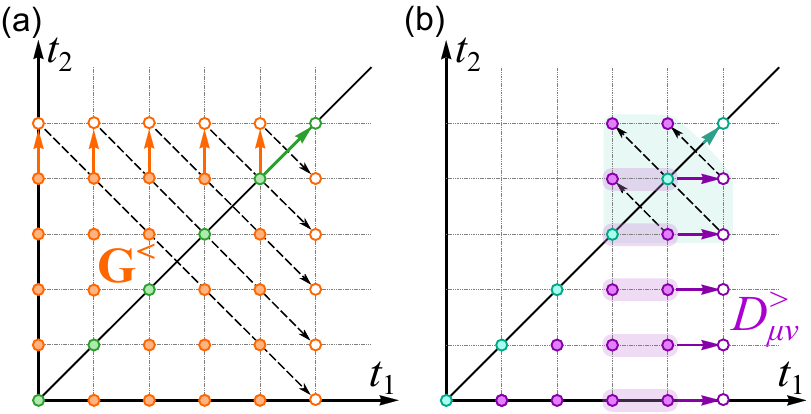}
  \caption{Propagation scheme for solving the KBEs
    eq.~\eqref{eq:KBEel} and \eqref{eq:KBEbos}. The time coordinates
    are discretized into a uniform mesh $\{t_n\}$, such that
    discrete-difference approximations can be applied to the
    derivatives.  (a) Solution scheme for the electron
    KBEs. $\mathbf{G}^<(t_{k},t_{n+1})$ is computed from
    \eqref{eq:KBEel2} (orange arrows). Analogously,
    $\mathbf{G}^>(t_{n},t_k)$ is obtained from eq.~\eqref{eq:KBEel1} 
    in the lower part of the time plane. Eq.~\eqref{eq:KBEdiag} 
    is used for propagating on the time diagonal (green
    arrows).  (b) Propagation method for the boson KBEs: $D^>_{\mu\nu}(t_{n+1},t_k)$,
    $k\le n$, is determined from $D^>_{\mu\nu}(t_{n-1},t_k)$ and
    $D^>_{\mu\nu}(t_{n},t_k)$, while the diagonal points are obtained from the surrounding
    grid points by a partial differential equation (see text). 
    Symmetry relations between of greater/lesser components are denoted by dashed arrows.
    \label{fig2}}
\end{figure}

The KBEs \eqref{eq:KBEel} for the electrons can be solved by standard
techniques. Specifically, we implemented a predictor-corrector Heun
method similar to
ref.~\onlinecite{stan_time_2009}. Eq.~\eqref{eq:KBEel1} is used for
propagating $\mathbf{G}^>(t_1,t_2)$ for $t_1>t_2$, while
$\mathbf{G}^<(t_1,t_2)$ is obtained from eq.~\eqref{eq:KBEel2} for
$t_1<t_2$ (see fig.~\ref{fig2}(a)). Eqs.~\eqref{eq:KBEel} are combined
at $t_1=t_2=t$ into the time-diagonal EOM
\begin{equation}
  \label{eq:KBEdiag}
  \iu \frac{\dd}{\dd t}\mathbf{G}^<(t,t) = \left[\mathbf{h}^\mathrm{MF}(t),
  \mathbf{G}^<(t,t)\right] + \Big(\mathbf{X}^<_L(t,t)+ \mathrm{h. c.}\Big) \ .
\end{equation}
For propagating the bosonic KBEs \eqref{eq:KBEbos} we have chosen the
Numerov method, as it provides a fourth-order scheme with minimal
number of function evaluations for this specific type of differential
equations. Generally, the method applies to
\begin{equation}
  \left(\frac{\dd^2}{\dd t^2}+W(t)\right) F(t) = S(t) \ ,
\end{equation}
which is transformed (after equidistant discretization $t_{n+1}-t_n =
\Delta t$) into the recursive relation
\begin{equation}
  \label{eq:numerov}
  \widetilde{F}_{n+1}-U_n \widetilde{F}_n + \widetilde{F}_{n-1} = 
  \frac{\Delta t^2}{12}\left(S_{n+1}+10 S_n + S_{n-1}\right)\ .
\end{equation}
Here, $S_n=S(t_n)$, $\widetilde{F}_n=(1-T_n)F(t_n)$, $U_n=(2+10
T_n)/(1-T_n)$ and $T_n=-(\Delta t^2/12) W(t_n)$. When applying the
Numerov method to the KBEs~\eqref{eq:KBEbos}, $S(t)$ plays the role of
the collision integral. From eq.~\eqref{eq:numerov} we see that
$S(t_{n+1})$ required to perform the step $t_{n}\rightarrow t_{n+1}$
is unknown at this point (similar to the Heun propagation
scheme). However, as $S(t_n)$ carries the dominant weight, we can
substitute $\left(S_{n+1}+10 S_n + S_{n-1}\right)\approx 12 S_n$ when
executing $t_n\rightarrow t_{n+1}$ for the first time. The precision
of eq.~\eqref{eq:numerov} is thereby reduced from fourth to second
order. Once the boson propagators are known up to $t_1,t_2\le
t_{n+1}$, the new collision integrals can be computed and the time
step $t_n \rightarrow t_{n+1}$ can be carried out in fourth order
according to eq.~\eqref{eq:numerov}. The analogous strategy applies to
eq.~\eqref{eq:EOMQre}.  We combine this corrector step with the one
needed for propagating the electron GF (as the electron (boson)
self-energy depends on the boson (electron) GF) and iterate until
self-consistency at each time step is achieved.

There is no need for computing $D^<_{\mu\nu}(t_1,t_2)$ for $t_1<t_2$
from eq.~\eqref{eq:KBEbos2}, as
$D^<_{\mu\nu}(t_1,t_2)=D^>_{\nu\mu}(t_2,t_1) =
-\big[D^>_{\mu\nu}(t_1,t_2)\big]^*$
(cf. appendix~\ref{app:basic}). Therefore, the propagation scheme can be
restricted to the lower time half-plane $t_2\le t_1$ for the greater
bosonic correlator (see fig.~\ref{fig2}(b)). At variance with the
electron KBEs it is not possible to formulate the time-diagonal EOM in
the form of eq.~\eqref{eq:KBEdiag}, as it relies on the notion of
first derivatives. We solve this issue by adding
eq.~\eqref{eq:KBEbos1} and \eqref{eq:KBEbos2} to obtain the
Poisson-type equation
\begin{align}
  \label{eq:KBEpois}
    -\frac{1}{\Omega_\mu+\Omega_\nu}&\left(\frac{\partial^2}{\partial t^2_1}
    +\frac{\partial^2}{\partial t^2_2}+\Omega^2_\mu+\Omega^2_\nu\right)D^>_{\mu\nu}(t_1,t_2) =
    Z^>_{\mu\nu}(t_1,t_2) \ ,\\ 
    (\Omega_\mu+\Omega_\nu)& Z^>_{\mu\nu}(t_1,t_2)
    =\Omega_\mu Y^>_{L,\mu\nu}(t_1,t_2) +\Omega_\nu Y^>_{R,\mu\nu}(t_1,t_2)  \ .
\end{align}
Next we apply the two-dimensional extension of the Numerov method (see
appendix~\ref{app:num2d}), expressing the Laplacian
$\nabla^2_{t_1,t_2} D^>_{\mu\nu}(t_n,t_n)$ by the nine surrounding
grid points $D^>_{\mu\nu}(t_{n+i},t_{n+j})$, $i,j=-1,0,1$. The
resulting equation can then be solved for
$D^>_{\mu\nu}(t_{n+1},t_{n+1})$ (sketched in fig.~\ref{fig2}(b)). Similar to the
one-dimensional case, the right-hand side of eq.~\eqref{eq:KBEpois}
has to be known at all these time points, as well, in order to achieve
fourth order. Analogously, we can approximate
$Z^>_{\mu\nu}(t_{n+i},t_{n+j})$ ($i,j=-1,0,1$) by
$Z^>_{\mu\nu}(t_{n},t_{n})$ when carrying out
$D^>_{\mu\nu}(t_{n},t_{n})\rightarrow D^>_{\mu\nu}(t_{n+1},t_{n+1})$
for the first time (predictor step) and apply several corrector steps
after computing $Z^>_{\mu\nu}(t_{n+1},t_{n+1})$. 

Collision integrals are computed by either Durand's rule (even number
of grid points) or Simpson's rule (odd number of points). The
momentum-momentum correlator required for calculating the boson
occupation eq.~\eqref{eq:occbos} is obtained from the mixed
derivative $\Omega_{\mu} \Omega_\nu D^{PP,>}_{\mu \nu}(t,t^\prime)=
\left[\partial_t \partial_{t^\prime}D^>_{\mu
    \nu}(t,t^\prime)\right]_{t=t^\prime}$.

\section{Application to photoemission from magnesium 2\lowercase{p}
  core state \label{sec:appl}}

In order to illustrate our propagation method for coupled
electron-boson Hamiltonians and, furthermore, explore the physics of
such systems in the time domain, we apply the theory developed in
sec.~\ref{sec:theory} to a typical process described by the s-model:
the photoemission from deep core state.

In particular we consider bulk Mg, a system where recent attosecond
streaking experiments \cite{lemell_real-time_2015} were able to
measure the time delay of photoemission between to the 2p core state
and the corresponding plasmonic satellite. The system is modeled by
the Hamiltonian \eqref{eq:Htot} with the static part
eq.~\eqref{eq:H0}. We account for the 2p state,
$\mathcal{E}_{i=2\mathrm{p}}$ and two virtual states
$\mathcal{E}_{i=k_{1,2}}$ representing photoelectrons. We consider one
boson mode (bulk plasmon) with energy $\Omega_\mathrm{pl} \simeq
10$~eV (subscript $\nu$ is dropped). For the electron-plasmon
interaction \eqref{eq:Hel-bos} we distinguish intrinsic ($\Gamma_\mathrm{in}$)
and extrinsic ($\mathbf{\Gamma}_\mathrm{ex}$) mechanisms:
\begin{equation}
  \label{eq:Hel-pl}
  \hat{H}_\mathrm{el-pl} = 
  \Gamma_\mathrm{in} \hat{c}_\mathrm{2p} \hat{c}^\dagger_\mathrm{2p} \hat{Q}
  + \sum_{i,j\ne \mathrm{2p}} \big(\mathbf{\Gamma}_\mathrm{ex}\big)_{ij}\,\hat{c}^\dagger_i
  \hat{c}_j \hat{Q} \ .
\end{equation}
The latter accommodate post-emission effects, that is inelastic
scattering of the emerging photoelectron from the electron sea upon
converting a part of their energy into a plasmon.
% When photoionizing the system with a laser with energy
% $\omega_\mathrm{L}$, it is expected to detect photoelectrons with
% energy
% $\mathcal{E}_{k_1}=\mathcal{E}_\mathrm{QP}+\omega_\mathrm{L}$
% (direct emission from QP peak $\mathcal{E}_\mathrm{QP}$,
% which is shifted with respect to $\mathcal{E}_{2\mathrm{p}}$ due to
% correlation) and
% $\mathcal{E}_{k_2}=\widetilde{\mathcal{E}}_\mathrm{2p}-\Omega_\mathrm{pl}+\omega_\mathrm{L}$
% (emission from PS). It is clear that both the intrinic and the
% extrinsic plasmon loss channel coalesce in the latter case. Hence, the
% dependence on $\Gamma_\mathrm{in}$ and $\mathbf{\Gamma}_\mathrm{ex}$
% needs to be studied on equal footing, especially since the
% photoemission signal is subject to quantum interference between the
% two pathways \cite{campbell_interference_2002}.

The s-model describes two phenomena: upon x-ray absorption, an excited
core electron looses a part of its energy and excites a plasmon. This
process is manifested as a sequence of PSs in the spectral function
occurring at lower energies. For fixed detector energy, larger photon
energy is needed to produce a photoelectron from PS as compared to QP.
In the reciprocal process the core hole is filled upon emitting an
x-ray photon. This process can be again accompanied by creating
plasmons, such that the electron looses a part of its energy and emits
a photon with smaller energy when recombining. For brevity we denote
PSs at lower (higher) energies as PS$^-$ (PS$^+$). Furthermore, the
position of the QP peak $\mathcal{E}_\mathrm{QP}$ is shifted by
$\Gamma^2_\mathrm{in}/2\Omega_\mathrm{pl}$ (correlation shift) \cite{Note1}
%
%\footnote{The factor $1/2$ when comparing to
%  ref.~\onlinecite{langreth_singularities_1970} is due to the
%  different definition of the electron-boson coupling
%  \eqref{eq:Hel-bos}.}
%
to larger energy with respect to the non-interacting value
$\mathcal{E}_\mathrm{2p}$ when the core state is occupied, whereas
$\mathcal{E}_\mathrm{QP}=\mathcal{E}_\mathrm{2p}-\Gamma^2_\mathrm{in}/2\Omega_\mathrm{pl}$
for the empty core state. The spectral functions for the two scenarios
are shown in fig.~\ref{fig4}(a).

\subsection{Time-dependent spectral function \label{subsec:tdafun}}

\begin{figure}[t]
  \centering
  \includegraphics[width=\columnwidth]{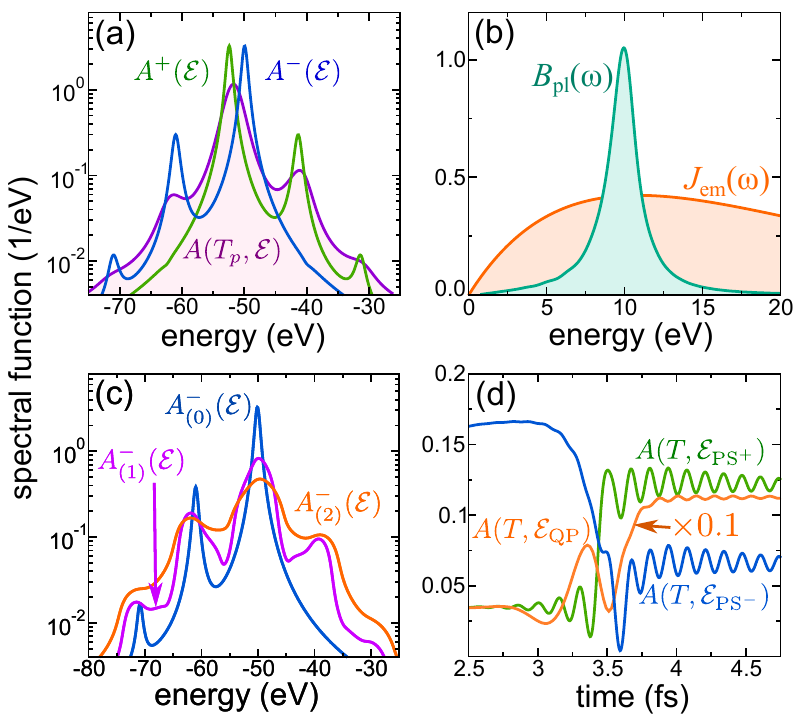}
  \caption{
    (a) 
    Spectral function $A^-(\mathcal{E})$
    ($A^+(\mathcal{E})$) of occupied (empty) core state, obtained by
    time propagation, along with $A(T_p,\mathcal{E})$.  
    (b) 
    Interacting plasmon spectral function $B_\mathrm{pl}(\omega)$ and,
    for comparison, the embedding density
    $J_\mathrm{em}(\omega)$. 
    (c)
    Electron spectral function $A^{-}_{(n)}(\mathcal{E})$
    (occupied core state) with fixed plasmon occupation
    $N_\mathrm{pl}=0,1,2$.
    (d)
    Dynamics of $A(T,\mathcal{E})$ for
    energy at the PSs and QP peak after the excitation,
    $\mathcal{E}=\mathcal{E}_{\mathrm{PS}^+}=-41.23$~eV,
    $\mathcal{E}=\mathcal{E}_{\mathrm{PS}^-}=-61.44$~eV, and
    $\mathcal{E}=\mathcal{E}_{\mathrm{QP}}=-51.76$~eV.
    \label{fig4}}
\end{figure}

It is now interesting to investigate the time-evolution in an
intermediate case, where the initially occupied 2p state is partially
photoionized, in real time. The spectral function is expected to
reorganize transiently, showing i) a shift of the QP peak and ii)
appearance of plasmonic satellites (PS$^+$). The energetic position of
these features in the spectral function varies in time primarily
reflecting changes in the core state occupation and in the number of
bosons in the system.

We solved the KBEs eq.~\eqref{eq:KBEel} and \eqref{eq:KBEbos} with the
algorithm from sec.~\ref{sec:numerics}. Instead of initializing with
the interacting Matsubara GFs, we switch on the interaction
adiabatically by defining
$s(t)=(1+\exp[\alpha(t_\mathrm{sw}-t)])^{-1}$ ($s(z)=0$ for $z \in
\mathcal{C}_\mathrm{im}$). Hence, initial correlations can be
disregarded ($\mathbf{\Sigma}^{\rceil,\lceil}=0$,
$\Pi^{\rceil,\lceil}_{\mu\nu}=0$), simplifying the propagation
scheme. We only consider intrinsic losses in this subsection, so
$\mathbf{\Gamma}_\mathrm{ex}=0$.

Plasmons typically decay by exciting particle-hole ($p\mhyphen h$)
pairs (Landau damping). Beside the states already incorporated in our
model, there might be other electronic transitions limiting the
plasmon lifetime. $p\mhyphen h$ excitations in the conduction band in
case of metals are a typical mechanism. In order to account for this
plasmon decay channel in a simple way, we add a bosonic bath. For the
latter we assume that the bath boson occupation number is zero, such
that we obtain
\begin{align}
  \label{eq:PIem}
  \Pi^{\mathrm{em,\gtrless}}(t_1,t_2) &= \sum_{\alpha}
  |\gamma_{\alpha}|^2 d^{\mathcal{B},\gtrless}_\alpha(t_1,t_2) = 
  -\frac{\iu}{2}\sum_\alpha |\gamma_{\alpha}|^2 e^{\mp \iu
  \omega_\alpha (t_1-t_2)} \nonumber\\
  &\equiv -\frac{\iu}{2} \int^\infty_{0}\!\dd \omega\, J_\mathrm{em}(\omega)  e^{\mp \iu
  \omega (t_1-t_2)}      \ ,
\end{align}
where $ J_\mathrm{em}(\omega)$ denotes the spectral density of the
bath (it includes the coupling). For a simple Ohmic bath \cite{weiss_quantum_2012} adopted
here eq.~\eqref{eq:PIem} can be analytically integrated:
\begin{align}
  J_\mathrm{em}(\omega)=g_0
  \frac{\omega}{\omega^2_\mathrm{c}}e^{-\omega/\omega_\mathrm{c}} \ , \
  \Pi^{\mathrm{em,\gtrless}}(t_1,t_2) = \frac{\iu g_0}{2(\omega_c
  (t_1-t_2)\mp \iu)^2}
\end{align}
The transition $\omega_\mathrm{c}\rightarrow \infty$ represents the
counterpart to the wide-band limit approximation (WBLA) often
encountered for the electron embedding self-energy, as
$\Pi^{\mathrm{em},\mathrm{R}}(t_1,t_2)\rightarrow -\pi
(g_0/2\omega^2_\mathrm{c})\delta'(t)$ , turning the EOM
\eqref{eq:EOMQre} for the boson amplitude into the equation for the
ordinary damped driven oscillator (similarly for the boson KBEs
\eqref{eq:KBEbos}), which has no memory. Adiabatic switching is
realized by $g_0 \rightarrow g_0 s(t_1)s(t_2)$.

We propagated the KBEs up to $T_p=15$~fs (time step $\Delta
t=0.024$~fs) with a switch-on time $t_\mathrm{sw}=5$~fs and
$\alpha=0.1$.  The inverse temperature is set to $\beta=50$~a.u.,
simulating the zero-temperature case. Initially, electronic levels are
occupied according to the Fermi function with chemical potential
$\mu=0$, while we assume for the plasmon occupation
$N_\mathrm{pl}(t=0)=0$. The environmental coupling leads to a non-zero
steady-state boson number as a result of the broadening of the
spectral function. For $g_0=1$~eV and $\omega_\mathrm{c}=10$~eV we
find $N_\mathrm{pl}=0.012$ for $t>t_\mathrm{sw}$. This is in
accordance with the thermodynamical equilibrium value obtained from
solving the Matsubara Dyson equation for the plasmon mode (including
embedding only). The boson spectral function, calculated analogously
to eq.~\eqref{eq:tdAfun}, is shown together with the Ohmic spectral
density of the bath in fig.~\ref{fig4}(b).

\begin{figure}[t]
  \centering
  \includegraphics[width=\columnwidth]{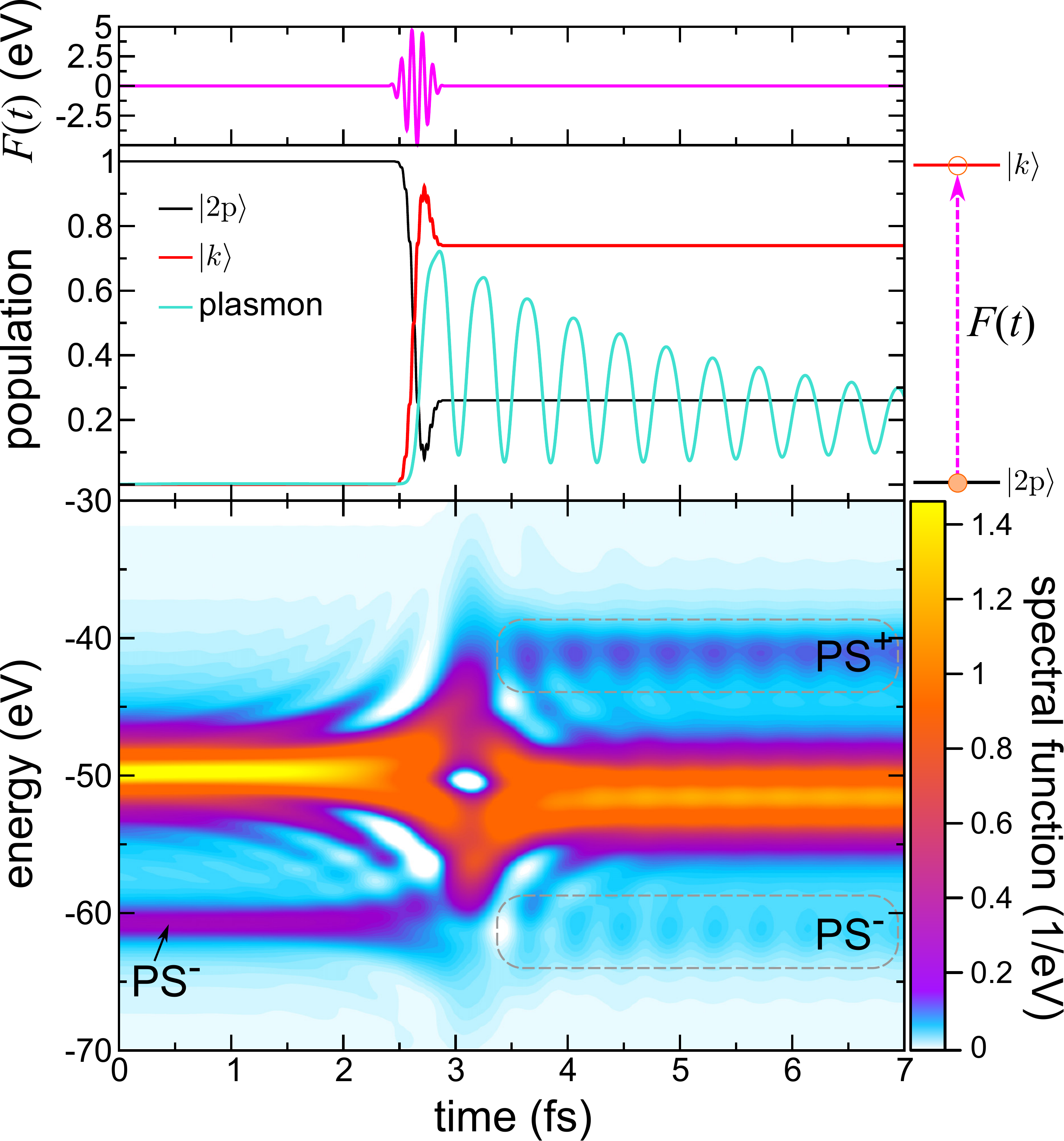}
  \caption{Top panel: laser matrix element $F(t)$. Middle panel:
    population dynamics of the 2p and the photoelectron state along
    with the plasmon occupation. Bottom panel: time-dependent spectral
    function $A(T,\mathcal{E})$ of the core state. The Fourier
    transformation eq.~\eqref{eq:tdAfun} has been slightly smoothed by
    including exponential damping.\label{fig5}}
\end{figure}

To simulate the ultra-fast photoionization dynamics, a laser pulse of
0.5~fs length and frequency $\omega_\mathrm{L}=55$~eV (see
fig.~\ref{fig5}, top panel) is applied after the system is fully
thermalized. For this quasi-resonant transition, we include one
conti\-nuum state $|k\rangle$ at
$\mathcal{E}_k=\mathcal{E}_\mathrm{QP}+\omega_\mathrm{L}$, where
$\mathcal{E}_\mathrm{QP}=-50$~eV is the QP energy for the no-hole
ground state. The light-matter interaction is simplified to
$F_{\mathrm{2p},k}(t)=F_{k,\mathrm{2p}}(t)\equiv
F(t)$. Electron-plasmon coupling is set to
$\Gamma_\mathrm{in}=5$~eV. We employ the self-energy
eq.~\eqref{eq:GWA1}.

Once the solution of the KBEs has been obtained, the time-resolved
spectral function can be computed by
\begin{align}
  \label{eq:tdAfun}
  \mathbf{A}(T,\mathcal{E})&=\iu \int\!\dd t\, e^{\iu\,\mathcal{E} t} 
  \Big[\mathbf{G}^>\left(T+\frac{t}{2},T-\frac{t}{2}\right)\nonumber \\
  &\quad\quad\quad-
   \mathbf{G}^<\left(T+\frac{t}{2},T-\frac{t}{2}\right) \Big] \ .
\end{align}
The laser-induced dynamics is presented in fig.~\ref{fig5}. The laser
pulse (fig.~\ref{fig5}, top panel) partially ionizes the core state
(the amplitude of $F(t)$ has been chosen to maximize the depopulation)
upon inducing plasmonic dynamics (fig.~\ref{fig5}, middle panel). The
creation of the core hole is faster than the plasmon time scale
$\tau_\mathrm{pl}=2\pi/\Omega_\mathrm{pl}$, indicating a strongly
non-adiabatic limit \cite{hedin_transition_1998} of intrinsic plasmon
excitation. The sudden change in the plasmon population is followed
then by oscillations in the plasmon occupation. This nonequilibrium
dynamics also becomes manifest in the time-resolved spectral function
$A(T,\mathcal{E})\equiv A_{\mathrm{2p},\mathrm{2p}}(T,\mathcal{E})$
(fig.~\ref{fig5}, bottom panel). The QP peak shifts transiently in
about 1.5~fs from the initial configuration (QP
peak at $\mathcal{E}=\mathcal{E}_{\mathrm{QP}}$, PS$^-$ at
$\mathcal{E}_{\mathrm{PS}^-}=60.3$~eV \cite{Note2}
%
%\footnote{Note the distance between QP peaks and PSs is overestimated
%  by the $GW$ approximation and slightly modified by the finite width
%  of the boson spectral function due to embedding.})
%
to the new QP position at $\mathcal{E}=-51.76$~eV. The shift is less
than expected from equilibrium spectral function for the completely
empty core state ($A^+(\mathcal{E})$, fig.~\ref{fig4}(a)). The
spectral density quenches transiently into the new equilibrium
position. In this way the PS$^-$ splits into a branch coalescing in the QP
and a second one merging with the shifted PS$^-$ after the pulse. A PS
above the QP appears (PS$^+$) as expected. Furthermore, the strength
of the PS$^\pm$ oscillates in time. In order to understand this
behavior one needs to take the bosonic occupation into account, as
well. Revisiting the equilibrium case, fig.~\ref{fig4}(c) depicts the
spectral function of the occupied core state with fixed integer
plasmon number $N_\mathrm{pl}=n$, $A^{-}_{(n)}(\mathcal{E})$. The
presence of a plasmon gives rise to a PS on the right-hand side of the
QP peak, describing plasmon-assisted photoemission (i.\,e., a plasmon
can be absorbed, transferring its energy to the
photoelectron). Increasing $n$ leads to stronger boson fluctuations
(cf. appendix~\ref{app:basic}) and hence enhances the magnitude of the
imaginary part of self-energy, leading to broadened spectral
features. This is consistent with the broadening observed in
fig.~\ref{fig4}(c). Returning to the time-dependent scenario, these
features are indeed manifested in $A(T,\mathcal{E})$
(fig.~\ref{fig5}): the spectral strength of the PS$^\pm$ displays
oscillations in phase with the time-dependent plasmon occupation
$N_\mathrm{pl}(t)$. Furthermore, the weight of the QP peak is
suppressed anti-phase-wise to the variations of the PSs weight, as
apparent from cuts of $A(T,\mathcal{E})$ at the characteristic
energies (fig.~\ref{fig4}(d)). Hence the spectral function exhibits an
oscillatory transfer of spectral weight from the QP peak to the
PSs. The enhanced broadening expected from fig.~\ref{fig4}(c) is
clearly visible in the spectral function at the end of the propagation
$A(T_p,\mathcal{E})$, as compared to the equilibrium spectra
$A^\pm(\mathcal{E})$ in fig.~\ref{fig4}(a).

\subsection{Time-resolved photoelectron spectra}

After analyzing the intrinsic effects upon removing the electron from
the core level, we proceed by incorporating extrinsic effects into the
electron-plasmon coupling \eqref{eq:Hel-pl}. Extrinsic
plasmon losses are post-emission, or, in other words, scattering
effects. This goes beyond the standard treatment of (time-resolved)
photoemission in terms of the lesser GF
restricted to bound states. Extrinsic effects can be incorporated by
explicitly including (at least) two states $|k_1\rangle$,
$|k_2\rangle$ representing photoelectrons and assign the plasmonic
matrix element $\Gamma_\mathrm{ex}\equiv
(\mathbf{\Gamma}_\mathrm{ex})_{k_1
  k_2}=(\mathbf{\Gamma}_\mathrm{ex})_{k_2 k_1}$ (diagonal elements are
set to zero). As photoelectrons propagate to infinity, the system is
treated as open in their subspace. This is accomplished by including
embedding self-energies. 
%
% Hereby we convey the following
% picture. Restricting the simulation region to the vicinity of the
% sample leads to a discretization of the continuum, such that only a
% finite number of continuum states $|k_j \rangle$ needs to be accounted
% for (corresponding to our states $|k_{1,2}\rangle$). On the other
% hand, $|k_j\rangle$ is not exact eigenstates of the SP Hamiltonian of
% the whole space, so that off-diagonal elements and thus coupling to
% the true continuum states (that form the bath in this case)
% occur. 
%
Because the continuum of photoelectron states describes
electron propagating outside the sample, a non-interacting basis can
be chosen. Defining the coupling density $U_{ij}(\mathcal{E})= \sum_k
V_{i,k}V^*_{k,j} \delta(\mathcal{E}-\epsilon_k)$, the continuum
embedding self-energy can be expressed in spectral representation as
\begin{equation}
  \mathbf{\Sigma}^{\mathrm{em},\gtrless}(t_1,t_2)=\int^\infty_0\!\frac{\dd \mathcal{E}}{2\pi} \,
  \mathbf{\Sigma}^{\mathrm{em},\gtrless}(\mathcal{E})e^{-\iu \mathcal{E} (t_1-t_2)} \ ,
\end{equation}
with 
\begin{align}
   \mathbf{\Sigma}^{\mathrm{em},<}(\mathcal{E})&= \iu\, \mathbf{U}(\mathcal{E}) 
   N_F(\mathcal{E}-\mu) \ , \\
  \mathbf{\Sigma}^{\mathrm{em},>}(\mathcal{E})&=-\iu\, \mathbf{U}(\mathcal{E})  
  \left[1-N_F(\mathcal{E}-\mu)\right] \ ,
\end{align}
where $N_F(\mathcal{E})$ denotes the Fermi distribution function.
$\mathbf{U}(\mathcal{E})$ accommodates the density of continuum states
(e.\,g. proportional to $\sqrt{\mathcal{E}}$ for free particles) and
matrix element effects. We simplify the expressions by approximating
$\mathbf{U}(\mathcal{E}) \approx U_0\mathbf{I}$ as a constant
(WBLA). The retarded embedding self-energy attains
$\mathbf{\Sigma}^{\mathrm{em,R}}(t_1,t_2)=-(\iu/2)U_0 \delta(t_1-t_2)$
in this case. In accordance with the physical picture we furthermore
assume that no electrons can return from the continuum, leading to
$\mathbf{\Sigma}^{\mathrm{em},<}(\mathcal{E}) \approx 0$.
The WBLA has an advantage that all states, regardless of their energy,
are damped uniformly. Such structureless embedding does not introduce
any additional energy dependent time delays.

The embedding self-energy for continuum states furthermore allows for
computing the photocurrent (the number of electrons emitted per unit
of time) $J(\mathcal{E},t) = \dd n(\mathcal{E},t)/\dd t$ by the
transient Meir-Wingreen formula
\cite{jauho_time-dependent_1994,myohanen_kadanoff-baym_2009} often
used in transport calculations. The Meir-Wingreen expression for the
total electron current flowing out of system reads
\begin{equation}
  \label{eq:MWF1}
  \frac{\dd n(t)}{\dd t} = 4 \mathrm{Re}
  \left\{\mathrm{Tr}_k\big[\mathbf{\Sigma}^{\mathrm{em},\mathrm{R}}\cdot \mathbf{G}^< + 
    \mathbf{\Sigma}^{\mathrm{em},\mathrm{<}}\cdot \mathbf{G}^\mathrm{A} +
    \mathbf{\Sigma}^{\mathrm{em},\rceil}\star \mathbf{G}^\lceil \big](t,t)
  \right\} \ .
\end{equation}
Here, $\mathrm{Tr}_k$ stands for the partial trace over the
photoelectron states $|k_{1,2}\rangle$. Eq.~\eqref{eq:MWF1} is
simplified in our case as $\mathbf{\Sigma}^{\mathrm{em},\rceil}=0$ due
to the adiabatic switching procedure and
$\mathbf{\Sigma}^{\mathrm{em},<}=0$ by the assumptions above. If we
further resolve with respect to the photoelectron energies $\mathcal{E}$, we
obtain
\begin{align} 
  J(\mathcal{E},t)&= 4 U_0\, \mathrm{Im}\!\int^t_0\!\dd \bar{t}\, e^{-\iu \mathcal{E}(t-\bar t)}
  \mathrm{Tr}_k\left[\mathbf{G}^<(\bar t,t)\right] \ .
\end{align}

We solved the KBEs eq.~\eqref{eq:KBEel} and \eqref{eq:KBEbos} for the
three-level system as in subsec.~\ref{subsec:tdafun} (with the
additional embedding self-energy included). In order to reflect the
experimental situation \cite{lemell_real-time_2015} the photoelectron
states are assigned energies $\mathcal{E}_{k_1} = 68$~eV and
$\mathcal{E}_{k_2} = 58$~eV
($\mathcal{E}_{k_1}-\mathcal{E}_{k_2}=\Omega_\mathrm{pl}$). The laser
frequency is chosen $\omega_\mathrm{L}=118$~eV, corresponding to the
transition $|2\mathrm{p} \rangle \rightarrow |k_1\rangle$. The pulse
length is set to $\tau_\mathrm{p}=1.2$~fs, corresponding to the full
width at half-maximum of 450 attoseconds as in the experiment. The
laser field amplitude is adjusted to perform a complete population
transfer in a non-interacting reference system.

\begin{figure}[t]
  \centering
  \includegraphics[width=\columnwidth]{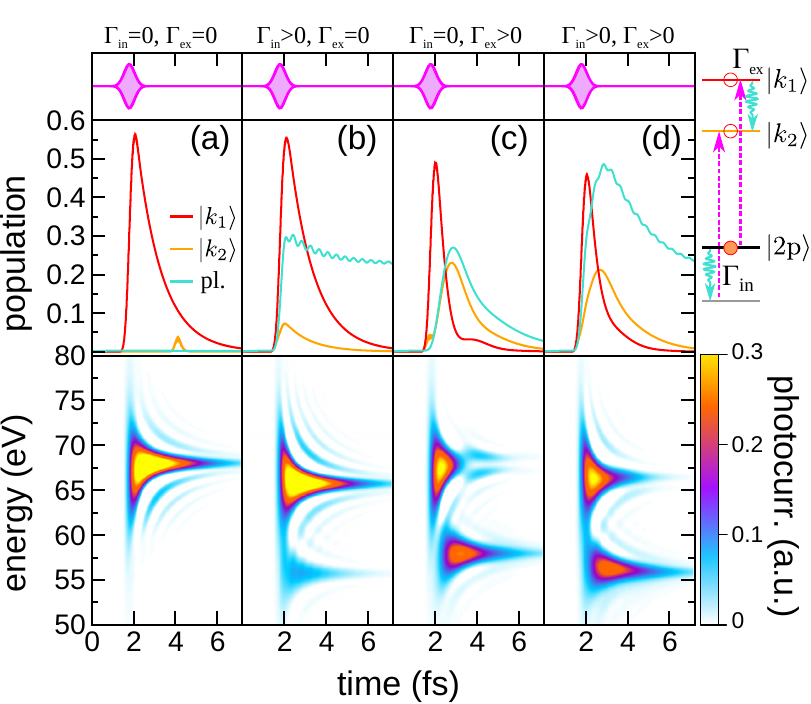}
  \caption{Laser-induced KBE dynamics for the three-level system (see
    sketch on the right-hand side). Top panels: envelop of the laser
    pulse. Middle panels: population dynamics of $|k_{1,2}\rangle$ and
    the plasmon occupation. Bottom panels: time-resolved photocurrent
    $J(\mathcal{E},t)$. The electron-plasmon interaction is set to (a)
    $\Gamma_\mathrm{in}=\Gamma_\mathrm{ex}=0$, (b)
    $\Gamma_\mathrm{in}=5$~eV, $\Gamma_\mathrm{ex}=0$, (c)
    $\Gamma_\mathrm{in}=0$, $\Gamma_\mathrm{ex}=1$~eV, and (d)
    $\Gamma_\mathrm{in}=5$~eV, $\Gamma_\mathrm{ex}=1$~eV. The
    magnitude of $\Gamma_\mathrm{ex}$ is chosen to put on a level
    with intrinsic/extrinsic effects. \label{fig6}}
\end{figure}

Both intrinsic and extrinsic losses can result in the population of
$|k_2\rangle$ continuum state with a lower energy, which translates to
a peak of $J(\mathcal{E},t)$ around $\mathcal{E} \sim 58$~eV (see
illustration in fig.~\ref{fig6}). The interplay between the two
channels can be studied by turning either $\Gamma_\mathrm{in}$ or
$\Gamma_\mathrm{ex}$ on or off (fig.~\ref{fig6}). In case that
electron-plasmon interaction is switched off completely, $|k_2\rangle$
acquires only a negligible occupation, while the transient
photoelectron spectrum $J(\mathcal{E},t)$ converges to a dominant peak
around $\mathcal{E} \sim 68$~eV (fig.~\ref{fig6}(a)). The plasmon
number $N_\mathrm{pl}$ stays, of course, constant. When including
intrinsic losses (fig.~\ref{fig6}(b)), $J(\mathcal{E},t)$ displays a
peak originating from the emission from PS$^-$. Note that the
$N_\mathrm{pl}(t)$ exhibits only small onset oscillations as compared
to fig.~\ref{fig5} because of the slower ionization process due to
the increased laser pulse duration. Pure extrinsic electron-plasmon
interaction (fig.~\ref{fig6}(c)) gives rise to similar spectral
features, the occurrence of a peak in the time-resolved spectra is
however delayed with respect to the intrinsic case. This is clear
since $|k_2\rangle$ can only be populated as a result of additional
scattering of $|k_1\rangle$ involving the creation of a plasmon. This transition
rate is set by i) $\Gamma_\mathrm{ex}$, ii) the plasmon frequency and
iii) the population of $|k_1\rangle$. Turning to the case of
comparable intrinsic and extrinsic losses (fig.~\ref{fig5}(d)) the
total number of photoelectrons detectable around $E \sim 56$~eV
increases beyond the previous cases. Interestingly, the bump in the
population of $|k_2\rangle$ due to extrinsic losses is less pronounced
as in fig.~\ref{fig6}(c) and delayed by $\sim 120$~as. On the other
hand, the photocurrent extends over a longer period of time before
approaching zero. Both factors are expected to influence the observed
streaking time delay \cite{lemell_real-time_2015}. Moreover, the
extrinsic process is weakened by the presence of intrinsic losses.
This can be seen from that the maximum in the population of
$|k_2\rangle$ is slightly less pronounced when comparing
fig.~\ref{fig6}(c) and (d). The intrinsic channel is already creating
a plasmon, which acts as driving against extrinsic losses. This is a
manifestation of quantum interference between intrinsic and extrinsic
losses, analogous to ref.~\onlinecite{campbell_interference_2002}. The
plasmon dynamics is quite different in the intrinsic and extrinsic
cases, as well. For intrinsic coupling only, the plasmon occupation
quickly rises and is weakly damped due to the boson embedding
self-energy. At variance, the plasmon creation is delayed in the
extrinsic case and the plasmon occupation vanishes rapidly. The
dynamics is governed here by a dynamical balance $|k_1\rangle
\rightarrow |k_2\rangle$ upon creating plasmons and by the inverse
process (visible in, e.\,g., the non-monotonic behavior of the
population of the upper continuum state for $t>2$~fs). As both
continuum states are subject to environmental coupling, the transition
$|k_2\rangle \rightarrow |k_1\rangle$ due to plasmon absorption leads
to an effective plasmon damping.

\section{Conclusions}
In this work we developed a formalism for simultaneous propagation of
the coupled fermionic and bosonic Kadanoff-Baym equations. The marked
feature of our scheme is the treatment of bosonic correlators: coupled
first-order equations of motions for $\langle \Delta
\hat{Q}_\mu(z)\Delta \hat{Q}_\nu(z')\rangle$, $\langle \Delta
\hat{Q}_\mu(z)\Delta \hat{P}_\nu(z')\rangle$, and $\langle \Delta
\hat{P}_\mu(z)\Delta \hat{P}_\nu(z')\rangle$ correlators are
reformulated as a second-order equation of motion for the
coordinate-coordinate correlator and is efficiently propagated using
the two-dimensional Numerov formula. The two-times nonequilibrium
Green's functions, which are the solutions of these equations,
completely describe the transient spectral density of both subsystems
and allow to obtain the time-resolved photoemission spectra. Several
competing scattering mechanisms result in a photoelectron arriving at
the detector with a lower energy and time delay as compared to the
unscattered one. In our calculations of the photoemission from the 2p
core state of bulk Mg we put apart scattering processes taking place
before the photon interaction with the material and after the
photoionization event on the electron's way to the detector. The time
delay between the unscattered and scattered electrons is determined by
the strength of electron-plasmon interaction which is the dominant
scattering mechanism for electrons excited by the XUV photons as in
the experiment \cite{lemell_real-time_2015}. Interference between
these two scattering pathways has already been theoretically predicted
to modify the spectral strength of plasmonic satellites. Here we
demonstrate that the interference has also profound impact on the time
dependence of the photoelectron current.

\begin{acknowledgments}
  This work is supported by the German Research Foundation (DFG)
  Collaborative Research Centre SFB 762 Functionality of Oxide
  Interfaces (J. B. and Y. P.) and Grant Number PA 1698/1-1 (M. S. and
  Y. P.).
\end{acknowledgments}

\appendix

\section{Source-field method for deriving the equations of motion}
\label{app:eom}

Let us assume that our system is initially prepared in some canonical
ensemble set by the inverse temperature $\beta$. Using
contour-evolution operator, we can express the expectation value of
any operator $\hat{O}$ by
\begin{equation}
  \label{eq:expectgen}
  \langle \hat O(z) \rangle =
  \frac{\mathrm{Tr}\left[\mathcal{T}\exp\left(-\iu\int_\mathcal{C}\dd\bar{z}\hat
        H(\bar z)\right) \hat O(z)
    \right]}{\mathrm{Tr}\left[\mathcal{T}\exp\left(
        -\iu\int_\mathcal{C}\dd\bar{z}\hat H(\bar z)\right)\right]}  \ .
\end{equation}
Since embedding contributions to the self-energy are additive and
straightforward to construct, we can focus on the electron-boson part,
described by the general Hamiltonian
\begin{align}
  \hat{H}_0(z) &= \sum_{ij} h_{ij}(z)\hat{c}^\dagger_i \hat{c}_j
  + \sum_{\nu}\sum_{ij}\Gamma^\nu_{ij}(z)\hat{c}^\dagger_i \hat{c}_j \hat{Q}_\nu
 \nonumber \\
  &\quad+ \sum_{\nu}\frac{\Omega_\nu(z)}{2}\left(\hat{P}^2_\nu+\hat{Q}^2_\nu\right) \ .
\end{align}
The Hamiltonian is now modified by adding time-dependent fields
coupled to the bosons, that is
\begin{equation}
  \hat{H}_\xi(z) = \hat{H}_0(z) + \sum_{\nu} \xi_\nu(z) \hat{Q}_\nu\ .
\end{equation}
The Heisenberg representation of all operators are now to be
understood with respect to $\hat{H}_\xi(z)$.  The way the source field
couples to the system, the bosonic propagators can directly be
obtained by
\begin{equation}
  D_{\mu\nu}(z_1,z_2) =\frac{\delta \langle
    \hat{Q}_\mu(z_1)\rangle}{\delta \xi_\nu (z_2)}
  \Big\vert_{\xi_\nu = 0} \ ,
\end{equation}
showing again that the boson GF describes the fluctuation of the
amplitude $\langle \hat Q_\nu(z_1)\rangle$. In the following, all
functional derivatives with respect to $\xi_\nu(z)$ are understood to
be taken at $\xi_\nu = 0$.

\subsection{Fermionic Green's function and self-energy}
\label{subsec:fermigf}

With the help of the Heisenberg EOM, we obtain the fermion GF 
\begin{equation}
  \label{eq:eomfermigf1}
    \left(\iu \frac{\partial}{\partial z_1} - \mathbf{h}(z_1)\right)
    \mathbf{G}(z_1,z_2) = \mathbf{I}\,\delta(z_1,z_2)+\sum_\nu
    \mathbf{\Gamma}^\nu(z_1)\, \mathbf{\Upsilon}^\nu(z_1,z^+_1,z_2) \ ,
\end{equation}
with the higher-order correlator $\Upsilon^\nu_{kj}(z_1,z_2,z_3) = -\iu\langle \mathcal{T}
\hat c_k(z_1) \hat Q_\nu(z_2) c^\dagger_j(z_3) \rangle $. The superscript $+$ denotes an
infinitesimal shift to later times. It is straightforward to see that it can obtained by
varying the fermionic GF with respect to the source field:
\begin{equation}
  \label{eq:relups1}
  \mathbf{\Upsilon}^\nu(z_1,z_3,z_2) = \iu 
  \frac{\delta \mathbf{G}(z_1,z_2)}{\delta \xi_\nu(z_3)} 
  +\langle \hat Q_\nu(z_3) \rangle \mathbf{G}(z_1,z_2) \ .
\end{equation}
Formally, we can identify the correlator term on the right-hand side
of eq.~\eqref{eq:eomfermigf1} with the self-energy term, i.~e.
\begin{align*}
  \sum_\nu \mathbf{\Gamma}^\nu(z_1) \mathbf{\Upsilon}^\nu(z_1,z^+_1,z_2) = \int_\mathcal{C}\dd
  z_3\, \mathbf{\Sigma}(z_1,z_3) \mathbf{G}(z_3,z_2) \ .
\end{align*}
In order to access the self-energy directly, we need the inverse
GF. In particular, let us define the right-inverse by
\begin{equation}
  \label{eq:unitygf}
   \int_\mathcal{C}\dd z_3\, \mathbf{G}(z_1,z_3) \overleftarrow{\mathbf{G}}^{-1}(z_3,z_2) = 
   \mathbf{I}\,\delta(z_1,z_2) \ .
\end{equation}
Multiplying by $ \overleftarrow{\mathbf{G}}^{-1}(z_3,z_2)$ and integrating
over $z_3$ yields for the self-energy
\begin{align*}
  \mathbf{\Sigma}(z_1,z_2) &= \sum_\nu\mathbf{\Gamma}^\nu(z_1)\!\! \int_\mathcal{C}\dd (z_3 z_4)\, 
  \mathbf{\Upsilon}^\nu(z_1,z_4,z_3) \overleftarrow{\mathbf{G}}^{-1}(z_3,z_2) \delta(z^+_1,z_4) \ .
\end{align*}
Eq.~\eqref{eq:relups1} suggests a separation of the self-energy into
two terms. For reasons that will become clear below, this distinction
amounts to splitting the self-energy into a mean-field and
correlation (c) part. The mean-field part gives
\begin{align*}
  \mathbf{\Sigma}^\mathrm{MF}(z_1,z_2) = \delta(z_1,z_2) 
  \sum_\nu \mathbf{\Gamma}^\nu(z_1) \langle \hat Q_\nu(z_1)\rangle  \ .
\end{align*}
The correlation part of the self-energy so far reads
\begin{align*}
  \mathbf{\Sigma}^\mathrm{c}(z_1,z_2)&=\iu \sum_\nu \mathbf{\Gamma}^\nu\int_\mathcal{C}\dd (z_3 z_4) 
  \frac{\delta \mathbf{G}(z_1,z_3)}{\delta \xi_\nu(z_4)} 
  \overleftarrow{\mathbf{G}}^{-1}_{nj}(z_3,z_2) \delta(z^+_1,z_4)    \ .
\end{align*}
Let us introduce the new source field
\begin{equation}
\zeta_{ab}(z)=\sum_\nu \Gamma^\nu_{ab}(z) \langle \hat Q_\nu(z)\rangle \ ,
\end{equation}
which is exactly the mean-field contribution to the fermionic
self-energy. This is analogous to the case of Hedin's equations, where
the original source field (coupling to the density) is replaced by the
total electronic energy. The reasons for the modification is to carry out
the variation of the fermionic GF with respect to fermionic quantities
only. Using the chain rule for functional derivatives, we obtain 
\begin{align*}
  \Sigma^\mathrm{c}_{ij}(z_1,z_2)&=\iu \sum_\nu\sum_{nk}\sum_{ab}
  \Gamma^\nu_{ik}(z_1)\int_\mathcal{C}\dd (z_3 z_4 z_5) 
  \frac{\delta G_{kn}(z_1,z_3)}{\delta \zeta_{ab}(z_5)} 
  \\ &\quad\quad \times \overleftarrow{G}^{-1}_{nj}(z_3,z_2) 
 \frac{\delta \zeta_{ab}(z_5)}{\delta \xi_\nu(z_4)}\delta(z^+_1,z_4)  \ .
\end{align*}
Invoking the functional variation analog of integration by parts, we can transfer the
variation with respect to $\zeta_{ab}(z_5)$ to the inverse GF. Noting further $\delta
\zeta_{ab}(z_5)/\delta \xi_\nu(z_4) = \Gamma^\mu_{ab}(z_5) D_{\mu\nu}(z_5,z_4)$ one arrives at
\begin{align*}
  \Sigma^\mathrm{c}_{ij}(z_1,z_2)&=\iu \sum_{\mu\nu} \sum_{nk}\sum_{ab}
  \!\!\int_\mathcal{C}\dd (z_3 z_5) \Gamma^\mu_{ab}(z_5)
   G_{kn}(z_1,z_3)\\ &\quad\quad \times \Lambda_{njab}(z_3,z_2;z_5)
 D_{\mu\nu}(z_5,z^+_1) \Gamma^\nu_{ik}(z_1) \ ,
\end{align*}
where we have introduced the three-point vertex
\begin{equation}
  \label{eq:vertexfun1}
  \Lambda_{abcd}(z_1,z_2;z_3) = -\frac{\delta\overleftarrow{G}^{-1}_{ab}(z_1,z_2)}
  {\delta \zeta_{cd}(z_3)} \ .
\end{equation}

\subsection{Vertex function and Bethe-Salpeter equation}

The definition of the vertex function eq.~\eqref{eq:vertexfun1} is
completely analogous to the derivation of Hedin's equations by the
source-field method. Beside this correspondence, treating the object
$ \Lambda_{abcd}(z_1,z_2;z_3)$ as the usual vertex function is
justified as it obeys the Bethe-Salpeter equation (BSE). In order to
show this property, we first realize that
\begin{align*}
  \overleftarrow{G}^{-1}_{ab}(z_1,z_2) &= \left(-\iu
    \frac{\overleftarrow{\partial}}{\partial z_1} \delta_{ab} - \zeta_{ab}(z_1) \right)\delta(z_1,z_2)
  \\ &\quad-\epsilon_a \delta_{ab} - \Sigma^\mathrm{c}_{ab}(z_1,z_2)
\end{align*}
and thus
\begin{equation}
  \label{eq:vertexfun2}
  \begin{split}
    \Lambda_{abcd}(z_1,z_2;z_3) &= \delta_{ac}\delta_{bd}\delta(z_1,z_2)\delta(z_1,z_3) 
    \\ &\quad+\frac{\delta \Sigma^\mathrm{c}_{ab}(z_1,z_2)}{\delta \zeta_{cd}(z_3)} \ .
  \end{split}
\end{equation}
Similar to the strategy above, we employ the chain rule for functional
variation for the second term in eq.~\eqref{eq:vertexfun2} to transform
\begin{align*}
  \frac{\delta \Sigma^\mathrm{c}_{ab}(z_1,z_2)}{\delta \zeta_{cd}(z_3)} 
  = \sum_{mn} \int_\mathcal{C}\dd(z_4 z_5) \frac{\delta \Sigma^\mathrm{c}_{ab}(z_1,z_2)}
  {\delta G_{mn}(z_4,z_5)} \frac{\delta G_{mn}(z_4,z_5)}{\delta \zeta_{cd}(z_3)} \ .
\end{align*}
As usual, we introduce the four-point kernel for BSE as
\begin{equation}
  \label{eq:bsekernel}
  K_{abcd}(z_1,z_2;z_3,z_4) = \frac{\delta \Sigma^\mathrm{c}_{ab}(z_1,z_2)}
  {\delta G_{cd}(z_3,z_4)} \ .
\end{equation}
Next, we would like to express the variation $\delta
G_{mn}(z_4,z_5)/\delta \zeta_{cd}(z_3)$ by the inverse GF in order to
close the equation. This can be achieved by inserting the unity
relation eq.~\eqref{eq:unitygf}. We thus obtain
\begin{align*}
  \frac{\delta \Sigma^\mathrm{c}_{ab}(z_1,z_2)}{\delta \zeta_{cd}(z_3)} &=
  \sum_{mn} \sum_p \int_\mathcal{C}\dd(z_4 z_5 z_6) K_{ab mn}(z_1,z_2;z_4,z_5) \\ &\quad\times
   \frac{\delta G_{mp}(z_4,z_6)}{\delta \zeta_{cd}(z_3)} \delta_{pn}  \delta(z_5,z_6)
 \\ &= \sum_{mn} \sum_{pq} \int_\mathcal{C}\dd(z_4 z_5 z_6 z_7) K_{ab mn}(z_1,z_2;z_4,z_5) \\ &\quad\times
 \frac{\delta G_{mp}(z_4,z_6)}{\delta \zeta_{cd}(z_3)}G_{pq}(z_5,z_7)
 \overleftarrow{G}^{-1}_{qn}(z_7,z_6) \ .
\end{align*}
Now we apply the variation to $\overleftarrow{G}^{-1}_{qn}(z_7,z_6)$,
which then amounts to $\Lambda_{qncd}(z_7,z_6;z_3)$. Finally, we obtain the BSE
\begin{equation}
  \label{eq:bse1}
  \begin{split}
    \Lambda_{abcd}(z_1,z_2;z_3) &= \delta_{ac}\delta_{bd}\delta(z_1,z_2)\delta(z_1,z_3) \\
    &\quad + \sum_{mn} \sum_{pq} \int_\mathcal{C}\dd(z_4 z_5 z_6 z_7) K_{ab mn}(z_1,z_2;z_4,z_5) \\
    &\times G_{mp}(z_4,z_6) G_{pq}(z_5,z_7)\Lambda_{qncd}(z_7,z_6;z_3) \ .
  \end{split}
\end{equation}

\subsection{Boson Green's function and polarization\label{subsec:appaboson}}
By differentiating the position-position correlator twice using EOM for
position~\eqref{eq:HeisEOMQ} and momentum~\eqref{eq:HeisEOMP} operators we arrive at the
second order differential equation
\begin{align*}
  -\frac{1}{\Omega_\mu(z_1)}\left(\frac{\partial^2}{\partial z^2_1}
    +\Omega^2_\mu(z_1)\right) D_{\mu\nu}(z_1,z_2) &= \delta_{\mu\nu}\delta(z_1,z_2)\\ &\quad+
  \mathrm{Tr}\left[
    \mathbf{\Gamma}^\mu(z_1) \frac{ \iu\,\delta \mathbf{G}(z_1,z^+_1)}{\delta \xi_\nu(z_2)} 
    \right] \ .
\end{align*}
Here, we omit the terms originating from the bath bosonic coordinates. They can be added
straightforwardly.  In order to close the EOM we introduce the irreducible polarization
\begin{equation}
  \label{eq:irrpol}
  P_{abcd}(z_1,z_2) = - \iu\frac{\delta G_{ab}(z_1,z^+_1)}{\delta \zeta_{cd}(z_2)} \ .
\end{equation}
Like for the fermion GF we introduce the (right) inverse boson GF according to
\begin{equation}
  \sum_{\xi}\int_\mathcal{C}\dd z_3\, D_{\mu\xi}(z_1,z_3)[\overleftarrow{D}^{-1}]_{\xi\nu}(z_3,z_2) =\delta_{\mu\nu} \delta(z_1,z_2) \ .
\end{equation}
Hence, we can express the polarization part of self-energy $\Pi^\mathrm{p}_{\mu\nu}(z_1,z_2)$
for the bosons, implicitly defined by
\begin{equation}
   \mathrm{Tr}\left[\mathbf{\Gamma}^\mu(z_1)\, 
     \iu\frac{\delta \mathbf{G}(z_1,z^+_1)}{\delta \xi_\nu(z_2)}
     \right]
  = \sum_\zeta\int_\mathcal{C}\dd z_3\, \Pi^\mathrm{p}_{\mu\zeta}(z_1,z_3) D_{\zeta\nu}(z_3,z_2) \ ,
\end{equation}
as
\begin{align*}
  \Pi^\mathrm{p}_{\mu\nu}(z_1,z_2) &=\mathrm{Tr}\left[ \mathbf{\Gamma}^\mu(z_1)\!\!\int_\mathcal{C}\dd
  z_3\frac{\iu \delta \mathbf{G}(z_1,z^+_1)}{\delta \xi_\zeta(z_3)}
  [\overleftarrow{D}^{-1}]_{\zeta\nu}(z_3,z_2) \right]\\
  &= \sum_{mn}\sum_{ij}\Gamma^\mu_{ij}(z_1) \int_\mathcal{C}\dd (z_3 z_4)
\frac{\iu \delta G_{ji}(z_1,z^+_1)}{\delta \zeta_{mn}(z_4)} 
\frac{\delta\zeta_{mn}(z_4)}{\delta\xi_\zeta(z_3)} \\
  &\quad \times [\overleftarrow{D}^{-1}]_{\zeta\nu}(z_3,z_2)
\end{align*}
and thus
\begin{equation}
  \Pi^\mathrm{p}_{\mu\nu}(z_1,z_2)=\sum_{mn}\sum_{ij}\Gamma^\mu_{ij}(z_1)
  \Gamma^\nu_{mn}(z_2) P_{jimn}(z_1,z_2) \ . 
\end{equation}
Finally, we investigate how the polarization $P_{abcd}(z_1,z_2)$ can
be correlated to the fermionic GF. For this purpose we invoke the rule
\begin{equation}
  \begin{split}
    \frac{\delta G_{ab}(z_1,z_2)}{\delta\zeta_{cd}(z_5)} &= -
    \sum_{pq} \int_\mathcal{C}\dd (z_3 z_4) G_{ap}(z_1,z_3)\frac{\delta \overleftarrow{G}^{-1}_{pq}(z_3,z_4)}
    {\delta\zeta_{cd}(z_5)}
    \\ &\quad \times G_{qb}(z_4,z_2) \ ,
  \end{split}
\end{equation}
from which we obtain
\begin{equation}
  \label{eq:polggg}
  \begin{split}
    P_{abcd}(z_1,z_2) &= -\iu \sum_{pq} \int_\mathcal{C}\dd (z_3 z_4)G_{ap}(z_1,z_3) G_{qb}(z_4,z^+_1)\\
    &\quad \times \Lambda_{pqcd}(z_3,z_4;z_2) \ .
  \end{split}
\end{equation}

\section{Basic properties of the boson propagator \label{app:basic}}

From the definition eq.~\eqref{eq:plasmongf1} one infers
$D_{\mu\nu}(z_1,z_2)=D_{\nu\mu}(z_2,z_1)$. This implies for the greater/lesser
Keldysh components
\begin{equation}
  \label{eq:DGLsymm}
  D^\gtrless_{\mu\nu}(t_1,t_2)= D^\lessgtr_{\nu\mu}(t_2,t_1) \ ,
\end{equation}
such that the retarded boson propagator becomes a real function:
\begin{equation}
  \begin{split}
    D_{\mu\nu}^\mathrm{R}(t_1,t_2)&=\theta(t_1-t_2) \left[D^>_{\mu\nu}(t_1,t_2)-D^<_{\mu\nu}(t_1,t_2)\right] \\
    &=2\theta(t_1-t_2)\mathrm{Re}\left[D^>_{\mu\nu}(t_1,t_2)\right] \ .
  \end{split}
\end{equation}
Similarly, one obtains
$D^\mathrm{A}_{\mu\nu}(t_1,t_2)=(D^\mathrm{R}_{\nu\mu}(t_2,t_1))^* =
D^\mathrm{R}_{\nu\mu}(t_2,t_1)$ for the advanced GF. For the Matsubara
component on the other hand,
$D^\mathrm{M}_{\mu\nu}(\tau_1-\tau_2)=D_{\mu\nu}(t_0-\iu \tau_1,t_0-\iu \tau_2)$, the symmetry 
\begin{equation}
  D^\mathrm{M}_{\mu\nu}(\tau) = D^\mathrm{M}_{\nu\mu}(-\tau)
\end{equation}
holds. Hence, as compared to fermions or bosons with a single peak in a spectral function
representing one QP, there is no discontinuity in the diagonal Matsubara function for the
transition $\tau=0^-$ to $\tau=0^+$.

In equilibrium one can, as usual, assume that the greater/lesser
propagators (and thus the retarded and advanced, as well) depend on
the time difference $t_1-t_2$ only. Therefore we can switch to
frequency space $D^\gtrless_\nu(\omega) = \int\dd t\,e^{\iu \omega
  t}D^\gtrless_\nu(t)$. The symmetry relation eq.~\eqref{eq:DGLsymm} implies
\begin{equation}
  D^\gtrless_{\mu\nu}(\omega) = D^\lessgtr_{\nu\mu}(-\omega) \ .
\end{equation}
The spectral function is obtained from
\begin{equation}
  B_{\mu\nu}(\omega) = \iu \left[D^>_{\mu\nu}(\omega) -D^<_{\mu\nu}(\omega) \right] \ ,
\end{equation}
which, in turn, allows to characterize the greater/lesser boson GF by
the fluctuation-dissipation theorem
\begin{subequations}
\begin{align}
  D^<_{\mu\nu}(\omega) &= -\iu N_\mathrm{B}(\omega)B_{\mu\nu}(\omega) \ , \\
  D^>_{\mu\nu}(\omega) &= -\iu \big(N_\mathrm{B}(\omega)+1\big)B_{\mu\nu}(\omega) \ ,
\end{align}
\end{subequations}
where $N_\mathrm{B}(\omega)$ is the Bose distribution function.  For illustration, let us
consider the non-interaction case. From the definition eq.~\eqref{eq:plasmongf1} the bare
boson GF follows as
\begin{align}
  \label{eq:dGLnonint}
  d^\gtrless_\nu(t_1,t_2) &= -\frac{\iu}{2} \left[(N_\nu +1) e^{\mp \iu
    \Omega_\nu (t_1-t_2)} + N_\nu e^{\pm \iu \Omega_\nu(t_1-t_2)} \right]  \\
&= \mp\frac12 \sin\left[\Omega_\nu(t_1-t_2)\right]-\iu 
\left(N_\nu+\frac12\right)\cos\left[\Omega_\nu(t_1-t_2)\right]\nonumber
\end{align}
(in thermal equilibrium $N_\nu = N_\mathrm{B}(\Omega_\nu)$) and
\begin{equation}
  d_\nu^\mathrm{R}(t_1,t_2)= -\theta(t_1-t_2)
  \sin\left[\Omega_\nu (t_1-t_2) \right] \ .
\end{equation}
Fourier transforming Eq.~\eqref{eq:dGLnonint} yields
\begin{equation}
  d^\gtrless_\nu(\omega) = -\iu \pi \left[
   (N_\nu +1) \delta(\omega\mp\Omega_\nu) +  N_\nu \delta(\omega \pm
   \Omega_\nu)\right] \ ,
\end{equation}
from which the spectral function follows as
\begin{equation}
  b_{\mu\nu}(\omega) = \pi \delta_{\mu\nu}\Big(\delta(\omega - \Omega_\nu) - \delta(\omega+\Omega_\nu) \Big) \ .
\end{equation}
Using the property $N_\mathrm{B}(-\omega)=-(N_\mathrm{B}(\omega)+1)$ the normalization of
the spectral function can be verified:
\begin{equation}
  \int^\infty_{-\infty} \frac{\dd \omega}{2\pi} \, N_\mathrm{B}(\omega)b_{\mu\nu}(\omega)
  = \delta_{\mu\nu}N_\mathrm{B}(\Omega_\nu) +\frac12 \ .
\end{equation}
The retarded GF  reads
\begin{equation}
  d^\mathrm{R}_\nu(\omega) = \frac{\Omega_\nu}{(\omega + \iu \eta)^2-\Omega^2_\nu} \ ,
\end{equation}
where $\eta$ is a positive infinitesimal.  By complex continuation one finds, analogously
to eq.~\eqref{eq:dGLnonint}, the non-interacting boson Matsubara function as
\begin{align}
  d^\mathrm{M}_\nu(\tau) &= -\frac{\iu}{2} \left[(N_\nu +1) e^{
    -\Omega_\nu |\tau|} + N_\nu e^{\Omega_\nu|\tau|} \right]  \\
&= \frac{\iu}{2} \sinh(\Omega_\nu |\tau|)-\iu 
\left(N_\nu+\frac12\right)\cosh(\Omega_\nu \tau) \ . \nonumber
\label{eq:dGLnonint}
\end{align}

\section{Two-dimensional Numerov formula \label{app:num2d}}

Consider the differential equation
\begin{equation}
  \left(\frac{\partial^2}{\partial t^2_1} 
+ \frac{\partial^2}{\partial t^2_2}\right) F(t_1,t_2) + S(t_1,t_2) = 0 \ ,
\end{equation}
which we would like to solve numerically on a uniform two-dimensional
mesh up to fourth order in the grid spacing $\Delta t$. This can be
achieved by applying the Numerov discretization method, which can be
summarized in a compact way by
\begin{equation}
  \label{eq:numerov2d}
  \begin{split}
    &\sum^1_{i=-1}\sum^1_{j=-1} \Big[(\Delta t)^2 (A_i B_j + A_j B_i)
    F(t_1+i \Delta t,t_2 +j \Delta t) \\ &\quad +
    (\Delta t)^4 B_i B_j S(t_1+i\Delta t,t_2+j\Delta t)
    \Big]=0 \ .
  \end{split}
\end{equation}
The coefficients are defined by
\begin{subequations}
  \begin{equation}
    A_{-1}=12 \ , \quad A_0 = -24 \ , \quad A_1 = 12 \ ,
  \end{equation}
   \begin{equation}
    B_{-1}=1 \ , \quad B_0 = 10 \ , \quad B_1 = 1 \ .
  \end{equation}
\end{subequations}
Solving eq.~\eqref{eq:numerov2d}, for instance, for $F(t_1+\Delta
t,t_2+\Delta t)$ yields a fourth-order forward recursion formula.

\vfill
%\bibliography{mylib}
%\bibliographystyle{apsrev4-1}


\end{document}